\documentclass[amsmath,amssymb,aps,eqsecnum]{revtex4}
\usepackage[dvips]{graphicx}
 \usepackage{bm,bbm}
\usepackage{epsfig}

\begin{document}
\title{A brief introduction to multipartite entanglement}
\author{Ingemar Bengtsson$^1$ and Karol {\.Z}yczkowski$^{2,3}$}

\affiliation{$^1$Fysikum, Stockholm University, Sweden}

\affiliation {$^2$Jagiellonian University, Cracow, Poland} 
\affiliation{$^3$Center for Theoretical Physics, 
  Polish Academy of Sciences Warsaw, Poland}

 \date{December 22, 2016}

\begin{abstract}
  A concise introduction to quantum entanglement 
  in multipartite systems is presented.
 We review entanglement of pure quantum states 
 of three--partite systems analyzing the classes
 of GHZ and W states and discussing the monogamy relations.
 Special attention is paid to equivalence with respect to 
 local unitaries 
 and stochastic local operations, 
invariants along these orbits,
momentum map and spectra of partial traces.
We discuss absolutely maximally entangled states
and their relation to quantum error correction codes.
An important case of a large number of parties is also analysed
and entanglement in spin systems is briefly reviewed.
\end{abstract}

%
\maketitle

\medskip
\begin{center}
{\small e-mail:   ingemar@physto.se \quad karol@cft.edu.pl}
\end{center}


\section{Introduction}

These notes are based on a new chapter written to the second edition
of our book  {\sl Geometry of Quantum States.
An introduction to Quantum Entanglement} \cite{BZ06}.
The book is written 
at the graduate level for a reader familiar with the principles of quantum mechanics.
It is targeted first of all for readers who 
do not read the mathematical literature everyday, but 
we hope that students of mathematics and of the information sciences will find it useful as well, since they also may wish to learn about 
quantum entanglement.

Individual chapters of the book are to a large extent
independent of each other. For instance, we hope
that the new chapter presented here
might become a source of information 
on recent developments on quantum entanglement
in multipartite systems  also for experts working in the field.
Therefore we  have compiled these notes, which aim to present
an introduction to the subject as well as an up to date
review on basic features of pure states entanglement of
multipartite systems.

All references to equations or the numbers of section refers
to the draft of the second edition of the book.
To give a reader a better orientation on the topics
covered we provide its contents in Appendix A.
The second edition of the book includes also 
a new chapter 12 on discrete structures in the Hilbert
space and several other new sections.

%

\section{How much is three larger than two?}
\label{sec:intmult}

In the simplest setup one discusses quantum entanglement 
for bipartite systems, described 
by states in a composite Hilbert space ${\cal H}_{AB}$.
Any product state $|\psi_A\rangle \otimes |\psi_B\rangle$ 
is separable, and any other pure state is entangled. 
These notions can be generalized for multipartite
systems in a natural way. A state of a system
consisting of three subsystems and belonging to 
the Hilbert space ${\cal H}_{ABC}={\cal H}_A\otimes  {\cal H}_B \otimes {\cal H}_C$
is (fully) separable if it has a product form containing three factors, 
$|\psi_A\rangle \otimes |\psi_B\rangle \otimes |\psi_C\rangle$. 
All other states are entangled. 
This seems to be a simple and rather innocent extension, so one is tempted to pose a
delicate question: Is there any huge qualitative difference between quantum 
entanglement in composite systems containing three or more subsystems and the known case
of bipartite systems?

The answer is `yes'. Recall that a general  pure 
state of two subsystems with $N$ levels each,

\begin{equation}
|\psi_{AB} \rangle =\sum_{i=1}^N \sum_{j=1}^N  \Gamma_{ij} \;
  |i_A\rangle \otimes |j_B\rangle \ , 
\end{equation}

\noindent can {\sl always} be written in the form 
\begin{equation}
|\psi_{AB}\rangle =  ( U_A \otimes U_B ) \sum_{i=1}^N  \sqrt{\lambda_i} \;
   |i_A\rangle \otimes |i_B\rangle  \ ,
\end{equation}

\noindent and its entanglement properties are characterized by its Schmidt vector 
$\vec{\lambda}$. But a {\sl general} tripartite pure state
\begin{equation}
|\psi_{ABC} \rangle =\sum_{i=1}^N \sum_{j=1}^N \sum_{k=1}^N  \Gamma_{ijk} \;
  |i_A\rangle \otimes |j_B\rangle \otimes  |k_C\rangle  
\label{three-tensor} 
\end{equation}
\noindent {\sl cannot} be written in the form 
\begin{equation}
|\psi_{ABC}\rangle =  ( U_A \otimes U_B \otimes U_C ) \sum_{i=1}^N  \sqrt{\lambda_i} \;
   |i_A\rangle \otimes |i_B\rangle \otimes |i_C\rangle \ .
\label{three-bis} 
\end{equation}
\noindent As the dimension counting argument at the end of Section 9.2 
reveals, such 
states are really very rare (although we will see that they have interesting properties). 
Turning to the mathematical literature in order to standardize the tensor $\Gamma_{ijk}$ 
in some way, we learn many interesting things  \cite{LMV00,KB09,NL08}, 
but there is no magical recipe that solves our 
problem. Not all algebraic notions developed for matrices
work equally fine for tensors. 

In short, multipartite entanglement is much more sophisticated than bipartite entanglement, 
and it has a rich phenomenology already for pure states. 
If one considers the number of parties in a quantum composite system,
then three is much more than two, four is more than three, and so it goes on. 
The issue of entanglement in multipartite quantum systems 
deserves therefore a chapter of its own.  
We will focus our attention on multi--qubit systems and on pure states, 
otherwise several chapters would be needed!

\section{Botany of states}
\label{sec:mult234}

\begin{figure} [htbp]
   \begin{center} \
 \includegraphics[width=4.68cm,angle=270]{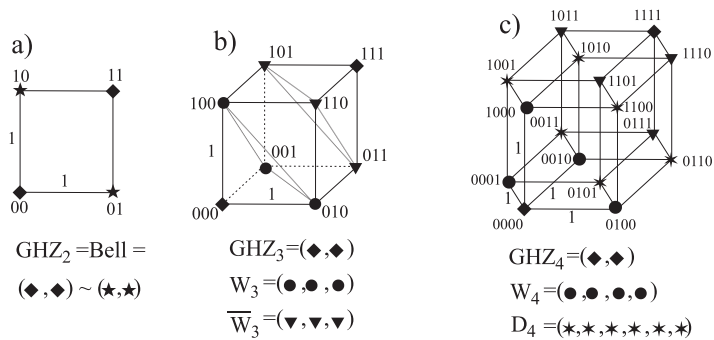}
\caption{Distinguished pure states for systems
of a) two, b) three and c) four qubits.
States represented by two corners at a diagonal of the square
represent the Bell states of two qubits. 
Two corners on a long diagonal of a cube represent
 the state $|GHZ_3\rangle = |000\rangle+|111\rangle$
while two points on a long diagonal of a hypercube
represent $|GHZ_4\rangle = |0000\rangle+|1111\rangle$.
Two locally equivalent states $W_3$ and ${\bar W}_3$ 
are formed by two parallel triangles (case b)
or tetrahedra ($W_4$ and ${\bar W}_4$, case c). 
$D_4$ denotes a four--qubit Dicke state 
(see Section \ref{sec:permutat}). 
}
\label{fig:cube4}
\end{center}
 \end{figure}

In Section 16.2 
we provided rather elaborate pictures showing where separable 
and entangled states can be found. If we are content with just a sampling of possibilities 
we can make do with less. Let the states in a product basis be represented by the corners 
of a square, in the fashion of Figure \ref{fig:cube4}a.
We use bitstrings to label the corners, so that $10$ (say) represents the state $|10\rangle 
\equiv |1_A\rangle \otimes |0_B\rangle$. Occasionally we will also use notation from the 
multipartite Heisenberg group, so that $Z = \sigma_z$, $X=\sigma_x$, $Y = iXZ = \sigma_y$. 
The qubit basis is chosen so that 
\begin{equation} Z|0\rangle = 
|0\rangle \ , \hspace{4mm} Z|1\rangle = -|1\rangle \ , \hspace{4mm} X|0\rangle = 
|1\rangle \ , \hspace{4mm} X|1\rangle = 
|0\rangle \ . \end{equation}
   
\noindent Any superposition of two states forming an edge of the square is separable. On 
the other hand, equal weight superpositions of states represented by two corners on 
a diagonal give maximally entangled Bell states, $|\phi^\pm\rangle = 
|00\rangle \pm |11\rangle$ and $|\psi^\pm \rangle = |01\rangle \pm |10\rangle$,
where normalization constants got suppressed.. 
Note that the local unitary 
transformation ${\mathbbm 1}\otimes Z$ interchanges $|\phi^+\rangle$ with 
$|\phi^-\rangle$ and  $|\psi^+\rangle$ with $|\psi^-\rangle$, 
while the equally local transformation ${\mathbbm 1}\otimes X$ interchanges 
the two diagonals of the square.  

This kind of picture is easily generalized to the case of three qubits. The eight 
separable basis states will form the corners of a unit cube, see 
Figure \ref{fig:cube4}b. To describe how far two corners are apart we count the 
number of edges one must traverse in order to go from the one to the other. 
The corners have been labelled so that this leads to the {\it Hamming distance} 
between two bitstrings of the same length,
equal to the minimal number of bits which need to be
flipped to transform one string into the other. 
\index{distance!Hamming} 
%
Superpositions of two states at Hamming distance one, 
belonging to the same edge of the cube, are separable. 
Superpositions of states at Hamming distance two  
display only bipartite entanglement. For instance, 
\begin{equation} |\psi_{AB|C}\rangle = 
\frac{1}{\sqrt{2}}(|000\rangle +|110\rangle ) = |\phi^+_{AB}\rangle 
\otimes |0_C\rangle \ . \end{equation}

\noindent States that cannot be decomposed in any such way are said to 
exhibit {\it genuine multipartite entanglement}.  

It is then thinkable that a 
balanced superposition of two states corresponding to two maximally distant
corners, at Hamming distance three, is highly entangled. This is indeed the case. The
{\it GHZ state}
\begin{equation}
|GHZ\rangle =\frac{1}{\sqrt{2}} \bigl( |000\rangle +|111\rangle \bigr) ,
\label{GHZ3}
\end{equation}

\noindent is named for Greenberger, Horne, and Zeilinger.
These 
authors went beyond Bell's theorem to obtain a contradiction between 
quantum mechanics and local hidden variables not relying on statistics 
\cite{GHZ89, GHSZ90}. 
The three--qubit GHZ state was created in 1999 \cite{BPDWZ99}.  
The present experimental record is a fourteen--qubit GHZ state 
in the form of trapped ions \cite{Monz11}.

Entanglement of the GHZ state is quite  fragile: 
If we trace out any subsystem 
from the GHZ state we obtain a separable state, which means that all the 
entanglement is of a global nature. 
Interestingly, this property holds if and only 
if the state is {\it Schmidt decomposable} \cite{Th99}, 
that is to say if it can be written on the form (\ref{three-bis}). Among such 
states the GHZ state is distinguished by the property that if one traces out 
any two subsystems, the maximally mixed state results. 

The GHZ state has many curious properties. If we rewrite it 
by introducing a new basis in Charlie's factor, $|+\rangle = |0\rangle + 
|1\rangle$ and $|-\rangle = |0\rangle - |1\rangle$, we find that it becomes 
\begin{equation} |GHZ\rangle = |\phi^+_{AB}\rangle \otimes |+\rangle_C + 
|\phi^-_{AB}\rangle \otimes |-\rangle_C \ , \end{equation}

\noindent where again the maximally entangled Bell states 
appear. At the outset all three parties agree that the state assignment 
governing the measurements of Alice and Bob is $\rho_{AB} = \mbox{Tr}_C|GHZ\rangle 
\langle GHZ|$. However, depending on what he chooses to do and on what the outcomes of 
his measurements are, Charlie can change his state assignment to one of $|0_A0_B\rangle$, 
$|1_A1_B\rangle$, $|\Phi_{AB}^+\rangle$, or $|\Phi_{AB}^-\rangle$. When Alice and 
Bob report the outcome of their measurement it will be consistent with that---as well 
as with the original state assignment $\rho_{AB}$. From this point of view the GHZ 
state describes entangled entanglement \cite{ZHG92}.

The GHZ state is an example of a stabilizer state, defined in Section 12.6
as an eigenstate of a maximal abelian subgroup of the 
three--partite Heisenberg group. Here this subgroup is generated by the four 
commuting group elements ${\mathbbm 1}ZZ, Z{\mathbbm 1}Z, ZZ{\mathbbm 1}, XXX$. 
All other equal weight superpositions related to two `antipodal'
corners of the cube separated by three Hamming units are locally equivalent to the 
GHZ state. There are eight of them, two on each diagonal 
(such as $|010\rangle \pm |101\rangle$), 
and all can be brought to the GHZ form by means of a local transformation 
(such as ${\mathbbm 1}X {\mathbbm 1}$) belonging to the Heisenberg group. 
And together these eight states form 
an orthonormal basis composed of states locally equivalent to the GHZ state. 

Another genuinely multipartite entangled state is the $W$ state
\begin{equation}
|W\rangle =\frac{1}{\sqrt{3}} \bigl( |001\rangle +|010\rangle + |100\rangle \bigr) .
 \label{W3}
\end{equation}
Some experts 
associate this name to the representation of the state in the space spanned by 
the energy and labels of the subsystems, which could resemble the letter $W$. 
Others tend to believe it is  related to the name of one of the authors of 
the paper \cite{DVC00} in which this state was investigated. 
It could be called `ZHG' or `anti--GHZ', as it 
appeared in a work \cite{ZHG92} by the authors of the 
GHZ paper \cite{GHZ89}. Incidentally, one can make a case for calling GHZ 
the `Svetlichny state' \cite{Sv87}.

\noindent In Figure \ref{fig:cube4}b it appears as a triangle. Its entanglement 
is more robust than that of the GHZ state, in the sense 
that an entangled mixed state results if we trace out any chosen subsystem, 
and indeed one can argue that in this respect its entanglement is maximally 
robust \cite{DVC00}.
Hence the $W$ state cannot be Schmidt decomposable. Indeed it 
cannot be written as a superposition of less than three
separable states \cite{DVC00}.

Introducing a fourth qubit into the game we need 
an extra dimension to construct a hypercube with $16$ corners. See Figure 
\ref{fig:cube4}c. Superpositions of two states at Hamming distance one 
are fully separable, as the tensor product consists of four factors.
Corners distant by two and three Hamming units 
correspond to states with bi- and tri-partite entanglement, respectively. 
Superpositions of two `antipodal' corners of the hypercube distant
by four Hamming units form genuinely four-party entangled states, such as 
\begin{equation}
|GHZ_4\rangle =\frac{1}{\sqrt{2}} \bigl(|0000\rangle +|1111\rangle \bigr) .
\label{GHZ4}
\end{equation}
This is called a four-qubit GHZ state, or sometimes 
a {\it cat state} 
to honour the Schr{\"o}dinger cat---which existed, or so we are told, in an 
equal weight superposition of classical states. 
A four qubit analogue of the W state corresponds to the
tetrahedron obtained by four permutations of the bitstring $0001$. Again the 
entanglement of the $W$ state is more robust than that of the GHZ state; we 
can define the {\it persistency} of entanglement \cite{BR01} as the minimum 
number of 
local measurements that need to be performed in order to ensure that the state is 
fully disentangled regardless of the measurement outcomes. For the GHZ$_K$ state 
the persistency equals 2, for the $W_K$ state it equals $K-1$. But this 
is not to say 
that the one state is more entangled than the other: they are entangled in 
different ways. 

The list of interesting states can be continued, but Figure \ref{fig:cube4} is 
already making it very clear that we are going to need an organizing principle to 
survey them. Botanists divide the Kingdom of Plants into classes, orders, families, 
genera, and species. It is reasonably clear that what corresponds to a division 
into species of the Kingdom of Multipartite States must be a division into orbits 
of the local unitary  
group. Hence we would say that two states $|\psi\rangle$ and $|\phi \rangle$ are 
{\it LU equivalent}, written $|\psi\rangle \sim_{\rm LU} 
|\phi\rangle$, if and only if there exist unitary operators $U_k$ such that  
\index{operations!LU}  
\begin{equation}
|\phi\rangle =  U_1\otimes U_2 \otimes \cdots \otimes U_K |\psi\rangle .
\label{LUKN}
\end{equation}
Since the overall phase of each unitary can be fixed,
we can choose unitary matrices with determinant set to unity and divide the set 
of states into orbits of the product group $G_U=SU(N)^{\otimes K}$, if we are 
studying $K$ quNits. 

Of more direct relevance to $K$ parties trying to perform some quantum task 
would be a division into {\it LOCC equivalent} sets of states, 
consisting of states that can be 
transformed (with certainty) into each other by means of local operations and classical communication, abbreviated LOCC. 
Fortunately, for pure states LU equivalence coincides 
with LOCC equivalence. In the case of two qubits 
this result was an immediate consequence of Nielsen's majorization theorem.
The argument in the multipartite case is similar but more involved.
For many  qubits this result is due to Bennett et al. \cite{BPRST00}. 
The criterion for SLOCC 
equivalence---which we are coming to---was presented by 
D\"ur, Vidal, and Cirac \cite{DVC00}. 

How much will this buy us? The state space of $K$ qubits has $2\cdot (2^K-1)$ real 
dimensions, while an orbit of the local unitary group has at most $3K$ dimensions 
because $SU(2)$ is a three dimensional group. Already for three qubits we can look 
forwards to a five dimensional set of inequivalent orbits.

A coarser classification may therefore be useful, corresponding to the division 
of plants into genera. This is offered by 
stochastic local operations and classical communication ({\it SLOCC}).
The definition of SLOCC operations presented 
in Section  16.4 
for the bipartite case
can be extended for multipartite systems in a natural way.
For completeness let us write an analogue of Eq. (16.44)
presenting a SLOCC transformation acting on a $K$--partite state,
\begin{equation} |\psi \rangle \ \rightarrow \ A_1\otimes A_2\otimes \dots 
A_K|\psi \rangle .
\label{SLOCCK} 
\end{equation} 
To formulate a relation 
between two SLOCC--equivalent states all matrices  have to be invertible,
\begin{equation}
|\psi\rangle \sim_{\rm SLOCC} 
|\phi\rangle =  L_1\otimes L_2 \otimes \cdots \otimes L_K |\psi\rangle.
\label{eqSLOCC}
\end{equation}

\noindent Since normalization does not play any role here we can freely assume that 
the matrices $L_i$ have unit determinant. 
Thus the group that governs SLOCC equivalence for a system consisting of $K$
subsystems with $N$ levels each 
is the special linear group composed with itself $K$ times, 
$G_L=SL(N,{\mathbb C}) ^{\otimes K}$.
 
The group $G_L$ is the complexification of the group $G_U$, and has twice as 
many real dimensions as the latter. Only twice, so again the number of 
inequivalent orbits grows quickly with the number of qubits. 
Like good botanists we must stand ready to introduce yet coarser classification 
schemes as we proceed. 
 
\section{Permutation symmetric states} 
\label{sec:permutat}

What we really need is a way of being able to recognize a given state at a glance 
(so that we do not have to rely on the way it looks like with respect to some 
perhaps arbitrary basis). We cannot quite do this, but we can if we restrict 
ourselves to the symmetric subspace ${\cal H}^{\otimes K}_{\rm sym}$ of the full 
Hilbert space ${\cal H}^{\otimes K}$. For $K$ qubits this subspace is in itself a 
Hilbert space of dimension $K+1$, and many states of interest such as the 
GHZ and $W$ states---as well as the ground states of many condensed matter systems, 
and more---reside in it. We will look at it through the glasses of the stellar 
representation from our Chapter 7.
Many nice descriptions of this idea can be found 
\cite{BKMGLS09,GKZ12,HKWGG09,Ma11,MKGLSB10}. 
Ours is perhaps closest to that of 
Ribeiro and Mosseri \cite{RM11}.

The symmetric subspace admits an orthonormal basis consisting of the states 
\begin{equation}
|K-k,k\rangle= {K \choose k}^{-\frac{1}{2}}\sum\limits_{\rm permutations} 
| 0\rangle^{\otimes K-k} \otimes | 1\rangle^{\otimes k},
\label{eq:dick}
\end{equation}
where $k=0,1,\dots, K$ and the summation is over all ${K \choose k}$ permutations 
of $K$--letter strings with $K-k$ symbols $|0\rangle$ and $k$ symbols $|1\rangle$. The 
basis states $|K-k,k\rangle$ can be identified with the Dicke states---which were thought of 
as angular momentum states in Chapter 7, 
but above all they were thought of 
as sets of stars placed at the North and South celestial poles only. 
The notion of composition of pure states, introduced in (7.27) 
 is useful here.
For any two symmetric states $|\psi_1\rangle$ and $|\psi_2\rangle$
of $K_1$ and $K_2$ qubits, respectively,  
the composite state $|\psi_1\rangle \odot |\psi_2\rangle$ of $K=K_1+K_2$ qubits 
is described by superposition of all $n$ stars. 
The reason why the stellar representation turns out to be useful is that  
the group of local unitaries acts on the symmetric subspace through its diagonal subgroup 
$SU(2)$. In other words, a local unitary transformation corresponds to a rotation of the 
celestial sphere housing the stars. Constellations of 
stars that can be brought into coincidence by means of a rotation correspond to states that are 
equivalent under local unitaries. 
%
\begin{figure}[htbp]
        \centerline{ \hbox{
                \epsfig{figure=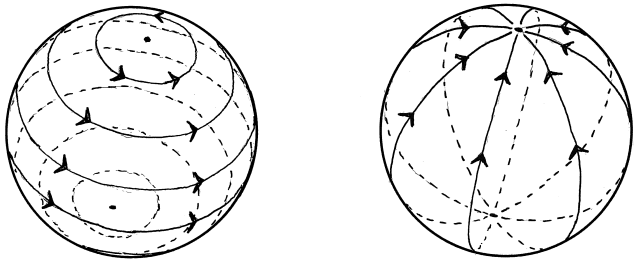, width=9.2cm}}}
        \caption{Left: A rotation of the celestial sphere, 
also known as an elliptic M\"obius transformation \cite{For51}. 
   Right: A Lorentz boost (a hyperbolic M\"obius transformation). 
 Combinations of the two are called four-screws 
\cite{PeRi84}. A general M\"obius transformation has two fixed points 
which can be placed anywhere on the sphere (4 parameters); the fixed points are
allowed to coincide. The amount of rotation and boost to include can also be 
chosen freely (2 parameters), so the full group has 6 dimensions. In our 
illustration the fixed points sit at the North and 
\index{transformation!M\"obius}
 South Poles (at 0 and $\infty$ on the complex plane). 
Constellations of stars that can be related by a M\"obius transformation represent SLOCC equivalent states.}
        \label{fig:multipartit1}
\end{figure}
A SLOCC transformation can also be visualized; it is effected by some 
$SL(2,{\mathbbm C})$ matrix. More precisely the group of M\"obius transformations 
acts effectively on the states, and this group is famously isomorphic to the 
group of Lorentz transformations $SO(3,1)$. 
\index{operations!SLOCC} 
\index{equivalence!SLOCC}
A general Lorentz transformations will cause the constellations of 
stars on the sky to change. In particular, if the observer moves with constant velocity towards 
the North Pole of the celestial sphere all the stars will be seen closer to the North Pole than 
they would be by an observer standing still.
This connection is explained 
by Penrose and Rindler \cite{PeRi84}. The best possible introduction to M\"obius transformations 
is the first chapter of the book by Ford \cite{For51}.
A detailed description of the action 
of the group is given in the caption of Figure \ref{fig:multipartit1}. 

Once these points are understood we can address the issue of how the symmetric 
subspace is divided into orbits under local unitaries and SLOCC
transformations. 
In fact the former question was already discussed, in considerable detail, 
in Section 7.2. 
The set of all symmetrized states has real dimension $2K$, and a typical orbit has dimension 3 
(equal to the dimension of the rotation group), so the set of all orbits has dimension $2K-3$ 
and will quickly become unmanageable as the number $n$ of qubits grows. For $K = 3$ things are still 
simple though. Any set of three stars sits on some circle on the sphere. By means of rotations 
this can be brought to a circle of constant latitude (one parameter), with one star at the 
Greenwich meridian, and two stars placed at other longitudes (two parameters). Hence the set of 
differently entangled three qubit states has three dimensions. By the time we get to four or more 
qubits a full description becomes at best very unwieldy. Still some special orbits are easily 
recognized in all cases. Although the physical interpretation is new, mathematically a GHZ 
state is identical to the noon state that we presented in 
Section 7.1, 
and its constellation of stars is a regular $K$-gon on some 
Great Circle on the sky. The set of all GHZ states form an orbit isomorphic to 
$SO(3)/Z_K ={\mathbbm R}{\bf P}^3/Z_K$. The $W$ states are Dicke states, and their stars are 
confined to two antipodal points on the sphere. The set of $W$ states is a 2-sphere.  

\begin{figure}[h]
\includegraphics[height=5.1cm]{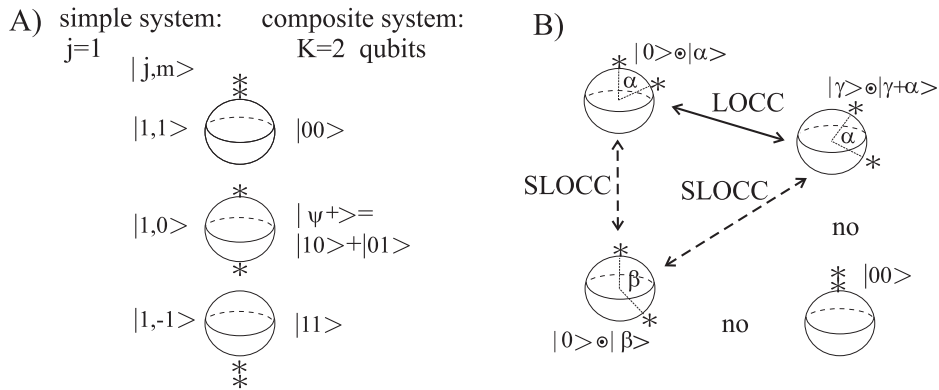}
\caption{Two stars on the sphere may represent
a state of a 3--level system or a symmetric
state of a two--qubit system.
a) Orthogonal basis $|j,m\rangle$ with $j=1$ and $m=-j,\dots,j$
is equivalent to $|00\rangle$, $|\psi^+\rangle$, $|11\rangle$.
b) Equivalence with respect to LOCC transformations 
(rotations of the sphere) and SLOCC transformations 
(preserving the degeneracy of the constellation).}
\label{fig:stars2}
\end{figure}

Things simplify if we consider states equivalent under SLOCC. For symmetric pairs of 
qubits there are only two orbits: one entangled, and one separable. 
See Figure \ref{fig:stars2}. 
Moving on to $K = 3$ qubits, and keeping 
in mind the picture of the effect a Lorentz boost has on the sphere, we see that 
every set of three non-coinciding stars can be brought into coincidence with three stars placed 
at the corners of a regular triangle on the equator. An equivalent way to see this is to observe 
that any three distinct complex numbers can be mapped to any other three by means of a M\"obius 
transformations. In fact there are only three orbits in this case. One orbit consists of all 
states with three coinciding stars. In Chapter 7 
these states were called $SU(2)$ coherent states. 
Now they reappear as (symmetric) separable states. Then there is an orbit consisting 
of states for which exactly two stars coincide. This is the W class. Finally there is the GHZ class 
for which all stars are distinct. This gives us three {\it degeneracy types}, denoted respectively 
by $\{ 3\}$, $\{ 2,1\}$, and $\{ 1,1,1\}$. Throwing randomly three stars 
on the sphere they will land in different points, hence type $\{ 1,1,1\}$ is the generic one. 
States with two stars merged together can be well approximated with states of the latter type, 
while the converse statement does not hold.

When $n = 4$ the classification remains manageable, but requires some work \cite{PeRi86,RM11}. We 
define the {\it anharmonic cross ratio} 
\index{invariants!cross ratio}
of four ordered complex numbers as 
\begin{equation} (z_1,z_2;z_3,z_4) = \frac{(z_1-z_3)(z_2-z_4)}{(z_2-z_3)(z_1-z_4)} \ . \end{equation}

\noindent Recall that we are on the extended complex plane, so that $\infty$ is a 
respectable number representing the South Pole of the Poincar\'e sphere. The point about 
this definition is that the cross ratio is 
invariant under M\"obius transformations, or in our language that this function of the positions 
of the stars is invariant under SLOCC. We are not quite done though, because we are interested 
in unordered constellations of stars, and as we permute the stars the cross ratio will assume 
six different values 
\begin{equation} \left\{ \lambda, \frac{1}{\lambda}, 1 - \lambda, \frac{1}{1-\lambda}, 
\frac{\lambda}{\lambda - 1}, \frac{\lambda-1}{\lambda} \right\} \ . \label{crgroup} \end{equation}

\begin{figure}[h]
        \centerline{ \hbox{
                \epsfig{figure=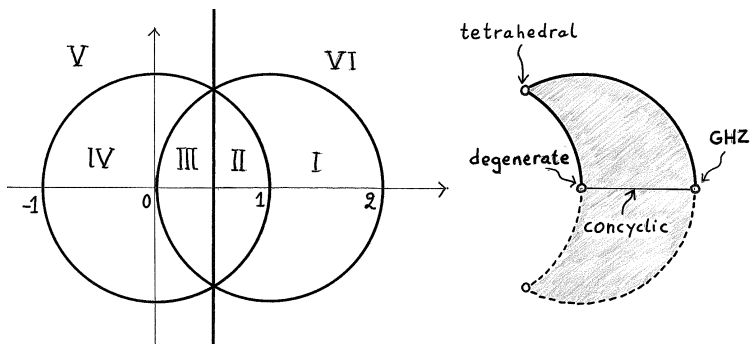, width=10.1cm}}}
        \caption{The cross ratio can take any value in the complex plane, but values related according 
to (\ref{crgroup}) are equivalent. Left: Therefore the complex plane is divided into 6 equivalent 
regions---the picture would look more symmetric if drawn on the Riemann sphere. Right: Using region 
I only we show the location of some states of special interest. Only 
a part of its boundary is included in the set of SLOCC equivalence classes.}
        \label{fig:multipartit2}
\end{figure}

\noindent The set of orbits is given by the set of values that the cross ratio can assume, provided 
it is understood that values related in this way represent the same orbit. 
%
%
This completely solves 
the problem of characterizing the set of differently entangled states of 4 symmetrized qubits, in 
the sense of SLOCC. Figure \ref{fig:multipartit2} is a map of the set of all SLOCC orbits in the 
4-qubit case. Note that if we pick two states at random they are likely to end up in different 
places on the map, and then they are inequivalent under SLOCC. 

Some interesting special cases deserve mention. 
First of all there are five different ways in which stars can coincide: $\{ 1,1,1,1\}$, 
$\{ 2,1,1\}$, $\{ 2,2\}$, $\{ 3,1\}$, and $\{ 4\}$. Thus there are five degeneracy types. 
Type $\{ 1,1,1,1\}$, in which all the 
stars sit at distinct points, is the generic one and the only one to contain a continuous family 
of SLOCC orbits. Within this family some special cases can be singled out. If the stars are lined 
up on a circle on the sky---they are then said to be {\it concyclic}---the cross ratio is real. 
If $\lambda = -1$, $2$, or $\frac{1}{2}$ 
the state is SLOCC equivalent to a GHZ state---the four stars form a regular 4-gon on a 
Great Circle. If $\lambda = 0$, $1$, $\infty$ two stars coincide. If $\lambda = e^{\pi i/3}$, 
$e^{-\pi i/3}$ the state is SLOCC equivalent to the {\it tetrahedral state}, 
\index{states!tetrahedral}
in which the stars form the vertices of a regular tetrahedron. 

The set of states for which some stars coincide is closed, but splits into 
four different orbits under SLOCC. Degeneracy type $\{ 2,1,1\}$ is the 
largest of these, but in itself it is an open set since states with more than two 
coinciding stars are not included. Its closure contains types $\{ 2,2\}$ and 
$\{ 3,1\}$, which are in themselves open orbits. Finally type $\{ 4\}$ consists 
of the separable states and forms a 2-sphere sitting in the closure of 
all the preceding types. 
Degeneracy types are called `Petrov types' 
by Penrose and Rindler \cite{PeRi86}.

As $K$ increases beyond 4 the story becomes increasingly complicated. It is not only 
difficult to survey the SLOCC orbits, it even becomes hard to count the degeneracy 
types---this is the number $p$ of partitions of the number $n$ into positive integers. 
This well known number theoretical problem was studied 
by Hardy and Ramanujan
 who obtained the asymptotic expression 
\begin{equation}
 p(K) \ \sim \  \frac{1}{4\sqrt{3}K} \exp \Bigl( \pi \sqrt{2K/3} \Bigr) , 
\label{eq:Raman}
\end{equation}
\noindent
valid when $K$ is large. 
The original result \cite{HR18}
goes back to 1918. An elementary proof of this formula
was given in 1942 by Erd{\"o}s \cite{Er42}
and simplified in 1962 by Newman \cite{Ne62}.

Distinguishing between the possible degeneracy types by means 
of suitable invariants is then practically impossible. A typical SLOCC orbit has dimension 
6 since this is the number of free parameters in 
the group $SL(2,{\mathbbm C})$, which means that the set of orbits representing states 
with $K$ non--coinciding stars always has real dimension $2K-6$. For $K=5$ we need two complex 
valued invariants---which are known functions of cross ratios, each cross ratio being a function 
of only four stars---but writing them down explicitly is a lengthy affair. For $K$ qubits 
we would need $K-3$ cross ratios, each computed using four stars at a time.  

In Section 7.4 
we mentioned the classical 
Thomson-problem: assume the stars are charged electrons, and place them
on a sphere in a way that minimizes the electric potential.
In a similar vein we can ask for the `maximally entangled' 
constellation. We can for instance try to maximize the Fubini-Study distance to the 
closest separable pure state. For $n \geq 3$ it is known that the latter lies in the 
symmetric subspace too (although the proof is not easy \cite{HKWGG09}). The actual 
calculation is hard; for $n = 3$ one finds that the $W$-state is the winner. For $n = 4$ 
numerical results point to the tetrahedral state as being `maximally entangled' in this 
sense, and for higher $n$ the resulting constellations tend to be spread out 
in interesting patterns on the sphere \cite{AMM10,BDGM15,Ma11,MGBBB10}.

The stellar representation was used here to describe symmetric
states of $K$ qubits. The same approach works also for symmetric states of a system
consisting of $K$ subsystems with $N$ levels each \cite{MRLL13}. Separable 
states are then $SU(N)$ coherent states, see Sections 6.4 and 6.5 
so a state can be represented by $K$ unordered points in ${\mathbbm C}{\bf P}^{N-1}$.
For  symmetric states of higher dimensional systems
a stellar representation works fine, but the sky
in which the stars shine is just larger than ours.

\section{Invariant theory and quantum states}
\label{sec:invarianc}

In this section we make a head-on attempt to divide the set of multi-qubit states 
into species and genera. 
We would like to have one or several simple functions of the 
components of their state vectors, such that they take the same values only 
if the states are locally equivalent. Let us call such functions {\it local 
unitary} or {\it SLOCC invariants} as the case may be. We also want enough of them, 
so that when we have evaluated them all we can say that the states are 
locally equivalent if and only if the invariants agree. 

In the case of bipartite entanglement we know how to do this, because the 
states $|\psi\rangle$ and $|\psi'\rangle$ are locally equivalent if and 
only if their Schmidt coefficients agree, that is if and only if the eigenvalues 
of the reduced density matrices $\rho_A = \mbox{Tr}_B|\psi\rangle \langle \psi|$ 
and $\rho'_A= \mbox{Tr}_B|\psi'\rangle \langle \psi'|$ agree. When we restrict 
ourselves to qubits only (with just one 
independent eigenvalue) this will happen if and only if their determinants  
agree. Letting the matrix $\Gamma$ carry the components of the state vector we 
see from Eq. (16.19) 
that 
\begin{equation} \det{\rho_A} = \frac{1}{N^2}|\det{\Gamma}|^2 \ , \end{equation}

\noindent For qubits, when $N = 2$, we are in fact dealing with the {\sl tangle}  
\begin{equation} \tau (|\Gamma \rangle ) = |\det{\Gamma}|^2 = 
\frac{1}{4}|\epsilon_{ij}\epsilon_{i'j'}\Gamma^{ii'}\Gamma^{jj'}|^2 \ . 
\label{tangledet} \end{equation}

\noindent From now  on we will be careful with tensors, just as we were in the 
earlier chapters. Because the transformation group is $U(2)\times U(2)$ there are 
two kinds of indices, unprimed and primed, and in the language of Section 1.4
 $\Gamma^{ij'}$ is a contravariant tensor.
In section 4.4  
 we used $A, B, \dots$ for the indices,
 but in this chapter $A$ is 
for Alice, $B$ is for Bob, and $C$ is (perhaps---opinions diverge on this point) 
for Charlie, so a change was called for. Indices always run from 0 to $N-1$ (and 
usually $N = 2$).
 
Indices may be summed over only if they are 
of the same kind and if one of them is upstairs and the other downstairs. Einstein's 
summation convention is in force, so repeated indices are automatically summed 
over. Thus when a local operation acts on a state it means that   
\begin{equation}
\Gamma^{ii'} \ \rightarrow \  \Gamma^{\prime \ ii'} = 
L^i_{ \ j} L^{i'}_{ \ j'}\Gamma^{jj'} \ , 
\label{matSLOCC}
\end{equation}

\noindent where $L^i_{\ j}$ and $L^{i'}_{\ j'}$ are two independent matrices, unitary 
or general invertible as the case may be. 
As usual the antisymmetric tensor $\epsilon_{ij}$ obeys $\epsilon_{00}=\epsilon_{11}=0$ 
and $\epsilon_{01}=1=-\epsilon_{10}$, always. This means that it transforms not like a 
tensor but like a tensor density, 
\begin{equation} \epsilon_{ij} \ \rightarrow \ \epsilon^{\prime}_{ij} = \epsilon_{kl}
L^k_{\ i}L^l_{\ j}(\det{L})^{-1} = \epsilon_{ij} \ . \end{equation}

\noindent Hence $\det{\Gamma}$ changes with a determinantal factor under general linear 
transformations. 
(See Schr\"odinger \cite{Sc50} for a short introduction to tensor 
calculus, and Penrose and Rindler \cite{PeRi84} for a long one.)
 This will become important as we proceed.

For a pair of qubits the tangle is the only 
invariant we need. Can we do something similar for many qubits? This kind of question 
is addressed, in great generality, in the field of mathematics called {\it invariant 
theory}. Invariant theory arose in the nineteenth century like Minerva ``from 
Cayley's Jovian head. Her Athens over which see ruled and which she served as 
a tutelary and beneficient goddess was {\sl projective geometry}.''
(The book by Olver \cite{Olv99} provides a very readable introduction to invariant theory; the 
book by Weyl \cite{Wey39} is a classic and 
we hope that our sample of his prose will attract the reader.)
 
\index{invariants}  
In its simplest guise the theory dealt with homogeneous polynomials in 
two variables, like those appearing in  Section 4.4, 
 and this is what we need for permutation invariant states. 
The simplest example is the quadratic form 
\begin{equation}
Q(u,v)  =  a_2 u^2 + 2 a_{1} uv + a_0 v^2 \ , 
\label{qutwo}
\end{equation}

\noindent (where we renamed the independent components of the symmetric 
multispinor as $a_2$, $a_1$, $a_0$). The expression is {\it homogeneous} in the 
sense that $Q(tu,tv)=t^2Q(u,v)$, hence it can be transformed to a second order 
polynomial $p(z) = Q(u,v)/u^2$ for the single variable $z = v/u$. Recall 
that what we really want to know is whether the polynomial $p(z)$ has multiple 
roots or not, and also recall that 
\begin{equation} p(z) = a_0z^2 + 2a_1z + a_2 = 0 \hspace{5mm} \Leftrightarrow 
\hspace{5mm} z = \frac{1}{a_0}(-a_1 \pm \sqrt{-\Delta}) \ , \end{equation}

\noindent where we defined the {\it discriminant} of the polynomial,
\begin{equation} \Delta \equiv a_0a_2 - a_1^2 \ . \end{equation}

\noindent The discriminant vanishes if and only if the roots coincide. Now suppose 
we make a M\"obius transformation, according to the recipe in Eq. (4.34). 
We declare that the quadratic form $Q$ transforms into a new quadratic form $Q'$, 
according to the prescription that 
\begin{equation}
Q'(u',v')  = Q'(\alpha u + \beta v, \gamma u + \delta v) = Q(u,v) \ . 
\label{qutwo2}
\end{equation}

\noindent This is a again a homogeneous polynomial of the same degree, and its 
coefficients $a'_i$ depend linearly on the original coefficients $a_i$. A small 
yet satisfying calculation confirms that 
\begin{equation} \Delta (a) = a_0a_2 -a_1^2 = (\alpha \delta - \beta \gamma )^2 
(a'_0a'_2 - a'^2_1) = (\det{G})^2\Delta (a') \ , \end{equation}

\noindent where $\det{G}$ is the non-zero determinant of the invertible two-by-two matrix 
occurring in Eq. (4.34). 
The calculation shows that the number of distinct 
roots will be preserved by any linear transformation of the variables, and it 
also makes the discriminant our---and everybody's---first example of a {\it relative 
invariant} under the group $GL(2,{\mathbbm C})$. 
\index{invariants!discriminant}
It is strictly invariant under $SL(2,{\mathbbm C})$.

This example can be generalized to homogeneous polynomials $Q(tu,tv) = t^nQ(u,v)$ of 
arbitrary order $n$, giving rise to $n$-th order polynomials in one variable, and 
indeed to homogeneous polynomials in any number of variables. Under a linear 
transformation of the variables Eq. (\ref{qutwo2}) will force the coefficients of $Q'$ 
to become linear functions of the coefficients of $Q$. An invariant of 
{\it weight} $m$ is a polynomial function of the coefficients 
which changes under invertible linear transformation only with a determinantal factor, 
\begin{equation} I(a)= (\det{G})^m I(a') \ . \end{equation}

\noindent We will also need {\it covariants} of weight $m$, which are polynomial 
functions of the coefficients and variables of a homogeneous polynomial $Q(u,v)$ 
\index{invariants!covariants}
such that 
\begin{equation} J(a,u,v) = (\det{G})^mJ(a',u',v') \ . \end{equation}

\noindent Evidently $Q$ itself is a covariant, but not the only one---fortunately, 
because we need enough of them, so that they characterize the behaviour of the 
$n$-th order polynomial $p(z) = Q(u,v)/t^n$. 

Now a polynomial $p(z)$ has a multiple root if and only if 
it has a root in common with the polynomial 
$\partial_zp(z)$. In general two polynomials, one of order $n$ and the other of order 
$m$, have a common root if and only if their {\it resultant} vanishes. By 
definition this is the determinant of a matrix which is constructed by writing the 
coefficients of the $n$-th order polynomial followed by $m-1$ zeros in the first row. In the 
next $m-1$ rows one shifts the entries of the previous row cyclically one step to the 
right. Then comes a row which contains the coefficients of the $m$th order polynomial 
followed by $n-1$ zeros, and this is followed by $n-1$ rows where the entries are 
permuted as before. 
(We present the construction as a piece of magic here, 
and refer to the literature for proofs \cite{Olv99}.)

We can now check if a cubic polynomial $p(z) = a_0z^3 + 3a_1z^2+3a_2z + a_3$ has a 
multiple root by taking the resultant with its derivative. It is convenient to divide 
by an overall factor, and define  
\begin{equation} \Delta = \frac{1}{27a_0}\left| \begin{array}{ccccc} 
a_0 & 3a_1 & 3a_2 & a_3 & 0 \\ 0 & a_0 & 3a_1 & 3a_2 & a_3 \\ 
3a_0 & 6a_1 & 3a_2 & 0 & 0 \\ 0 & 3a_0 & 6a_1 & 3a_2 & 0 \\ 
0 & 0 & 3a_0 & 6a_1 & 3a_2 \end{array} \right| \ .  \end{equation}

\noindent This invariant is called the discriminant of the cubic, and has 
weight 6 in the sense that $\Delta (a) = (\det{G})^6\Delta (a')$. It 
vanishes if and only if the corresponding quantum state has a pair of 
coinciding stars (in the sense of Section \ref{sec:permutat}). But it does 
not allow us to distinguish  between the 
degeneracy classes $\{ 2,1,1\}$ and $\{ 3\}$. To do so we need to bring in a covariant 
as well. For the cubic form we have the {\it Hessian covariant} $H$, which by definition is 
the determinant of the matrix of second derivatives of the form $Q$. In itself this is a 
quadratic form whose associated discriminant is precisely equal to $\Delta$. One can 
show that $H = 0$ and $Q \neq 0$ if and only the cubic polynomial has a triple root.  
An additional covariant, denoted $T$, is often listed. It obeys 
\begin{equation} T^2 = 2^43^6\Delta Q^2 - H^3 \ . \label{syzygy} \end{equation}

\noindent This is an example of a {\it syzygy} between invariants. (In astronomy 
a syzygy is said to occur when three celestial bodies line up on a straight line. The 
term is also used in poetry, but we omit the details.)

Note that the order of the discriminant, considered as a polynomial function 
of the coefficients, increases with the order of the polynomial. However, something 
interesting happens when we consider quartic polynomials (meaning, for us, symmetric 
states of four qubits). After calculating the discriminant $\Delta$ of the quartic 
according to the recipe, one can show that 
\begin{equation} \Delta = I_1^3 - 27I_2^2 \ , \end{equation}

\noindent where 
\begin{equation} I_1 = a_0a_4 - 4a_1a_3 + 3a_2^2 \ , \hspace{6mm} I_2 = 
\left| \begin{array}{ccc} a_0 & a_1 & a_2 \\ a_1 & a_2 & a_3 \\ a_2 & a_3 & a_4 \end{array} 
\right| \ . \end{equation}

\noindent Both $I_1$ and $I_2$ are in themselves invariants, of weight 4 and 6 respectively. 
Hence we have three invariants, but the third is given as a polynomial function of the two 
others. The interpretation 
is that the quartic has a multiple root if and only if $\Delta = 0$. It has a triple or 
quadruple root if and only if $I_1 = I_2 = 0$. The corresponding quantum state is SLOCC equivalent 
to the tetrahedral state if and only if $I_1 = 0$ and $I_2 \neq 0$. There are three covariants 
in addition to the two independent invariants. 

The difficulties increase with the order of the polynomial.
Even Sylvester 
got himself into trouble with the septic polynomial  
\cite{Olv99}, and a modern mathematician (Dixmier) says that 
``{\sl la situation pour $n \geq 9$ est obscure}''.
 We leave that as an exercise in French.

The question, how many independent polynomial invariants may exist occupied numerous 
eminent mathematicians. An important issue was to check whether there exists a finite 
collection of invariants, such that any invariant can be written as their polynomial 
function. A deep theorem was proved by Hilbert 
himself: 

\smallskip
\noindent {\bf Hilbert's basis theorem}. {\sl For any system of homogeneous 
polynomials, every invariant is a polynomial function of a finite 
number among them.} 
\smallskip

\noindent In high brow language the invariants belong to a finitely generated ring. In 
a way the theorem offers more information than we really need, since we may be happy 
to regard a set of invariants as complete once every other invariant can be given as 
some not necessarily polynomial function of that set, as happens in Eq. (\ref{syzygy}). 
Hilbert eventually provided a constructive proof of his theorem, but it is a 
moot question if his procedure can be implemented computationally. And we need a 
generalization of the original theorem, since we are interested in subgroups 
of the full linear group. Fortunately the result holds also for the subgroups 
we are interested in, namely the local groups $G_U$ and $G_L$ \cite{Olv99}. It 
also holds for some finite subgroups: an accessible example of what Hilbert's theorem 
is about is the fact that every symmetric polynomial can be written 
as a polynomial function of the elementary symmetric functions (2.11).
The 14-th problem on Hilbert's famous list
concerns the issue of finiteness of a polynomial invariant basis
for {\sl every} subgroup of linear invertible transformations.
 It was settled (in the  negative) in 1959 \cite{Gray00}. 

Leaving binary forms behind---or, in quantum mechanical terms, stepping out of the 
symmetric subspace---we can now try to characterize a given orbit of LU equivalent 
states by looking for the  local unitary invariants. Each individual invariant will 
be a homogeneous polynomial in the components of the state vector, and we start 
our search with a guarantee that there exists a finite number of independent 
invariants, in terms of which all other invariants can be obtained by taking sums 
and products. 
\index{invariants!local unitary}
If we can find such a set we are done:
to verify if two pure states are locally equivalent
it will be enough to compute a finite number of invariants for both states 
and check whether the corresponding values coincide.

But we still have to identify a suitable set of independent invariants.
In general this task requires lengthy calculations, and here we confine ourselves 
to describing the results in the three qubit case.
We rely on  Sudbery \cite{Su00} and others \cite{LPS99,BL01}. 
In their turn these authors relied on Weyl \cite{Wey39}.

Any 
invariant under the group $G_U$ of local unitaries  
can be represented as a homogeneous polynomial 
in the entries of the tensor (\ref{three-tensor}) and their complex conjugates. 
There exists only one invariant of order two, 
$I_1=\langle \psi | \psi \rangle = \Gamma^{i_1i_2i_3}{\bar{\Gamma}_{i_1i_2i_3}}$,
equal to the norm of the state. (We now need three different kinds of indices!) 
There are three easily identified polynomial invariants of order four,
\begin{eqnarray}
I_2 &=& {\rm Tr} \rho_A^2=
  \Gamma^{i_1i_2i_3} \Gamma^{j_1j_2j_3} {\bar \Gamma_{j_1i_2i_3}} {\bar \Gamma_{i_1j_2j_3}}, \\
I_3 & =& {\rm Tr} \rho_B^2=
  \Gamma^{i_1i_2i_3} \Gamma^{j_1j_2j_3} {\bar \Gamma_{i_1j_2i_3}} {\bar \Gamma_{j_1i_2j_3}}, \\
I_4 & =& {\rm Tr} \rho_C^2=
  \Gamma^{i_1i_2i_3} \Gamma^{j_1j_2j_3} {\bar \Gamma_{i_1i_2j_3}} {\bar \Gamma_{j_1j_2i_3}} .
\label{I_123}
\end{eqnarray}
These are the purities of the three single--party reductions.
By construction one has $1/2 \le I_i \le 1$  for $i=2,3,4$. Note that the use of 
Einstein's summation convention saved us from writing out six sums per invariant. 

At order six we have invariants such as ${\rm Tr}\rho_A^3$. However, because of the 
characteristic equation (see Section 8.1)   
applied to two-by-two matrices 
it happens that we have the syzygy 
$2{\rm Tr}\rho_A^3 = 3{\rm Tr}\rho_A{\rm Tr}\rho_A^2- ({\rm Tr}\rho_A)^3$, 
so this is an example of an invariant dependent on those we introduced 
already. An independent polynomial invariant of order six,  in this context called the  
{\it Kempe invariant} \cite{Kem99}, is 
\begin{equation}
I_5 =
  \Gamma^{i_1i_2i_3} \Gamma^{j_1j_2j_3} \Gamma^{k_1k_2k_3} {\bar \Gamma_{i_1j_2k_3}} 
{\bar  \Gamma_{j_1k_2i_3}} {\bar \Gamma_{k_1i_2j_3}} \ . \label{I_5}
\end{equation}

\noindent It is clearly symmetric with respect to exchange of the subsystems, and 
one can show that $2/9 \le I_5 \le 1$. The minimum is attained by the $W$ state. By 
considering various bipartite splittings of our systems one finds some 
further possibilities, but they are not independent invariants because 
\begin{eqnarray} I_5 = 3 {\rm Tr}[(\rho_A \otimes \rho_B)\; \rho_{AB}] 
- {\rm Tr}\rho_A^3  -{\rm Tr}\rho_B^3 = \hspace{55mm} \nonumber \\ \\ 
= 3 {\rm Tr}[(\rho_A \otimes \rho_C)\; \rho_{AC}] 
- {\rm Tr}\rho_A^3  -{\rm Tr}\rho_C^3 = 3 {\rm Tr}[(\rho_B \otimes \rho_C)\; \rho_{BC}] 
- {\rm Tr}\rho_B^3  -{\rm Tr}\rho_C^3 \ . \nonumber 
\end{eqnarray}

\noindent Readers who try to verify these identities should be aware that doing so 
requires very deft handling of the characteristic equation for two-by-two matrices 
\cite{Su00}. 

A sixth independent invariant is of order eight,
\begin{equation}
\label{I6}
 I_6 = |2{\rm Det}_3(\Gamma)|^2 \ ,   \end{equation}

\noindent where the {\it hyperdeterminant} of the tensor $\Gamma^{i_1i_2i_3}$ is, for 
the moment, defined to be
\begin{eqnarray} {\rm Det}_3(\Gamma) = \frac{1}{2}
\epsilon_{i_1j_1} \epsilon_{i_2j_2}\epsilon_{k_1\ell_1}
\epsilon_{k_2\ell_2}\epsilon_{i_3k_3}\epsilon_{j_3\ell_3}
 \Gamma^{i_1i_2i_3}\Gamma^{j_1j_2j_3}\Gamma^{k_1k_2k_3}\Gamma^{\ell_1 \ell_2\ell_3} = 
\hspace{10mm} \nonumber \\
\label{det33} \\ 
 =  [\Gamma_{000}^2\Gamma_{111}^2 + \Gamma_{001}^2\Gamma_{110}^2 
 + \Gamma_{010}^2\Gamma_{101}^2  + \Gamma_{100}^2\Gamma_{011}^2] \hspace{31mm}  \nonumber \\
   - 2 \bigl[ \Gamma_{000}\Gamma_{111} ( \Gamma_{011} \Gamma_{100}+ \Gamma_{101} 
	\Gamma_{010} +\Gamma_{110} \Gamma_{001} ) \hspace{29mm} \nonumber \\ 
  + \Gamma_{011}\Gamma_{100} ( \Gamma_{101} \Gamma_{010}+ \Gamma_{110} \Gamma_{001} ) +                           
  \Gamma_{101}\Gamma_{010}\Gamma_{110} \Gamma_{001} \bigr] \hspace{10mm} \nonumber \\
       +  4 \bigl[  \Gamma_{000} \Gamma_{110} \Gamma_{101} \Gamma_{011} + 
	\Gamma_{111}\Gamma_{001}\Gamma_{010} \Gamma_{100} \bigr]. \hspace{24mm} 
			\nonumber 
\end{eqnarray}
(Hyperdeterminants were introduced by Cayley in 1845 \cite{Ca45}, who had 
finalized the theory of determinants a couple of years earlier. Hyperdeterminants returned 
in force 150 years later in the book by Gelfand, Kapranov, and Zelevinsky \cite{GKZ94}, 
and were brought into this story by Coffman, Kundu, and Wootters \cite{CKW00}.)

\begin{figure}[h]
        \centerline{ \hbox{
         \epsfig{figure=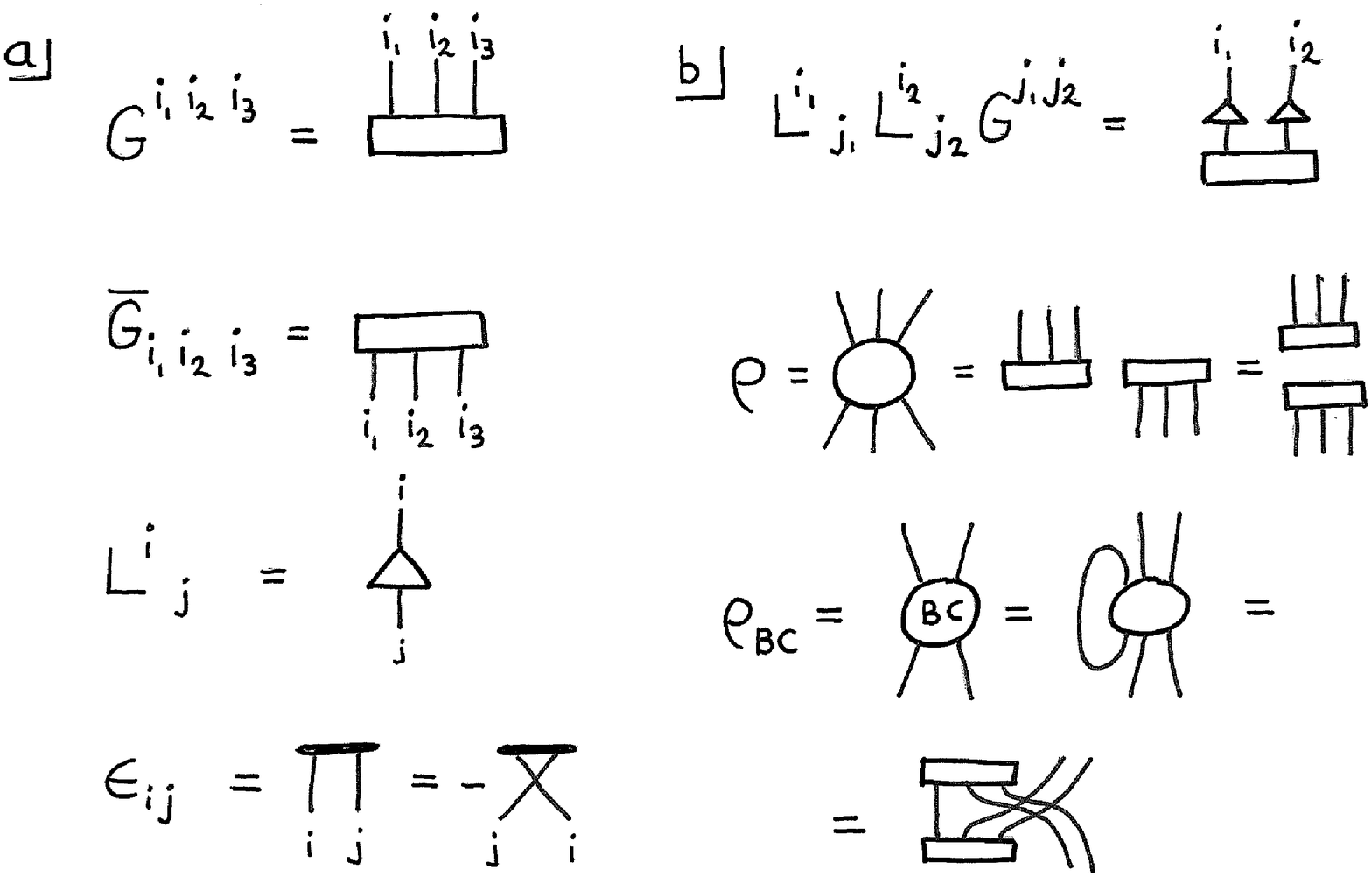, width=100mm}}}
        \caption{In graphical notation a tensor is represented 
as a shape with arms (upper indices) and legs (lower indices) attached. 
Open lines are labelled by the free indices, while lines 
connecting two tensors represent contracted indices (which need no label). 
A complication is that our tensors are tensors under a product group, so 
three different kinds of lines occur. In private calculations clarity is 
gained (but speed lost) if one uses three different colours to distinguish them. 
a) Four examples of tensors. b) New tensors from old. The pure state density 
matrix $\rho$ equals an outer product of two tensors; it 
does not matter how we place the latter on the paper. We also show $\rho_{BC}$, 
obtained from a $3$-party state $\rho_{ABC}$ by means of a partial trace.}
        \label{fig:birdtracks}
\end{figure}

\begin{figure}
        \centerline{ \hbox{
                \epsfig{figure=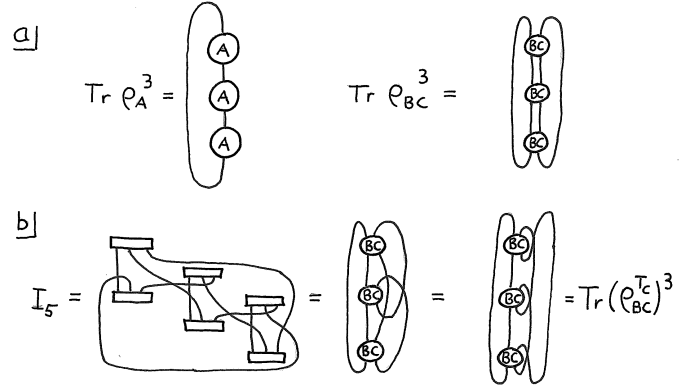, width=90mm}}}
  \caption{Examples of tensor calculations. 
  a) Here we take the trace of products of two 
different reduced density matrices. b) As a non-trivial exercise we prove 
that $I_5 = \mbox{Tr}(\rho_{BC}^{T_C})^3$ \cite{BTL12}, where $\rho_{BC}^{\rm T_C}$ 
is the partial transpose with respect to subsystem $C$ of the reduced density 
matrix Tr$_A\rho$.}
        \label{fig:birdtracks2}
\end{figure}

\noindent In the final expression indices are lowered for typographical reasons. 
To write out all the sums---and in fact for doing tensor algebra too---the graphical 
technique given in Figures \ref{fig:birdtracks}-\ref{fig:birdtracks2} is 
helpful. 
It is due to Penrose \cite{Pe71,PeRi84}, who also provided the 
mathematical rigour behind it. Graphical notation as such goes back to Clifford 
\cite{Olv99}. Coecke and his collaborators have devised graphical 
techniques tailored to quantum mechanics \cite{Coe06}.

It is clear that the hyperdeterminant is somewhat analogous to the ordinary determinant, 
\index{invariants!hyperdeterminant} 
and that the invariant $I_6$ is analogous to the single invariant (\ref{tangledet}) used in 
the two-qubit case. We postpone the general definition of hyperdeterminants 
a little, and just observe that each term is a product of 
four components such that their `barycentres' coincide with the centre of the cube
shown in Figure \ref{fig:cube4}b.
Examining the explicit expression for the hyperdeterminant 
one can convince oneself that $0 \le I_6 \le 1/4$. See also Table \ref{tab:3qubLU}.
\begin{table}[h]
\caption{LU invariants for some exemplary (normalized) states.}
 \smallskip
\hskip -0.5cm
{\renewcommand{\arraystretch}{1.21}
\begin{tabular}
[c]{cccccc} 
\hline \hline
State &  $I_2$  & $I_3$ & $I_4$  & $I_5$ & $I_6$ \\
  \hline \hline
                                separable  & $1$ & $1$ & $1$ &  $1$ & $0$ \\
$|\phi_A\rangle \otimes |\psi_{BC}\rangle $ & $1$ & $1/2$ & $1/2$ &  $1/4$ & $0$ \\
$|\phi_B\rangle \otimes |\psi_{AC}\rangle $ & $1/2$ & $1$ & $1/2$ &  $1/4$ & $0$ \\
$|\phi_C\rangle \otimes |\psi_{AB}\rangle $ & $1/2$ & $1/2$ & $1$ &  $1/4$ & $0$ \\
                               $|W\rangle $ & $5/9$ & $5/9$ & $5/9$ &  $2/9$ & $0$ \\
                             $|GHZ\rangle $ & $1/2$ & $1/2$ & $1/2$ &  $1/4$ & $1/4$ \\
\hline \hline
\end{tabular}
}
\label{tab:3qubLU}
\end{table}

If an invariant takes different values for two states then these states 
cannot be locally equivalent. Reasoning in the
opposite direction is more difficult, since we have to show that we have 
a complete set of invariants.
For a normalized state $|\psi_{ABC}\rangle$ the first invariant is fixed, $I_1=1$,
so we have only five remaining invariants to describe a local orbit.
Note that one needs at least five such numbers, as the
dimensionality of   ${\mathbbm C}{\bf P}^7$ ---
the space of three--qubit pure states --- is $14$
and the number of parameters of the local unitary
group  $SU(2)^{\otimes 3}$ is $3\times 3=9$.
Unfortunately our invariants are still not enough to uniquely single out a local orbit. 
Clearly $I_i(|\psi\rangle ) = I_i(|\psi\rangle^*)$ for all six invariants, so they 
cannot distinguish between a state and its complex conjugate. Grassl has found an 
additional invariant of order 12 which does complete the set \cite{AAJ01}. The 
somewhat more modest problem of deciding when two given multiqubit states can 
be connected by local unitaries has been discussed by Kraus \cite{Kra10}.  
\begin{figure} [htbp]
   \begin{center} \
 \includegraphics[width=4.34cm,angle=270]{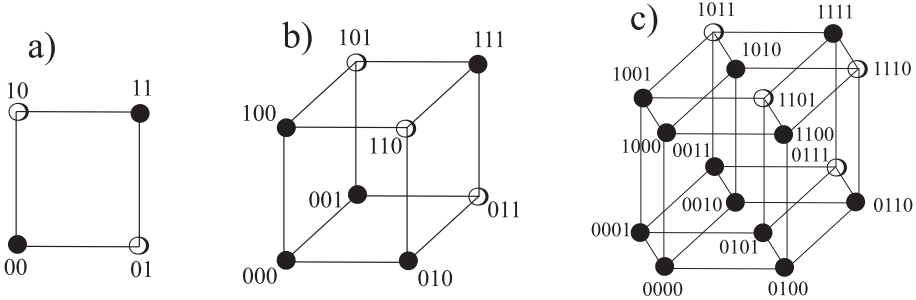}
\caption{Using the nearest separable state as one element of the basis, 
basis vectors at Hamming distance 1 from this state do not contribute to the 
superposition, as shown here with open circles for a) 2, b) 3, and c) 4 qubits.}
\label{fig:cube4b}
\end{center}
 \end{figure}

Having described the local orbits in terms of polynomial invariants
we can travel along the orbit in order
to find some state distinguished by a simple {\it canonical form}. 
In the two--qubit case the Schmidt decomposition provides an obvious choice. 
In the three--qubit case things are not so simple. By means of three local 
unitary transformation we can always set 3 out of the 8 components of a 
given state vector to zero. See Figure \ref{fig:cube4b}. (For $K$ quNits we can 
set $K\cdot N(N-1)/2$ components to zero \cite{CHS00}.)  
An elegant way to achieve this is to observe that there 
must exist a closest separable state, in the sense of Fubini-Study distance. 
By changing the bases among the qubits we can assume this to be the 
state $|111\rangle$. An easy exercise 
 (Problem 17.4) 
shows that this forces three of the components 
to be zero. Having come so far we can adjust the phase factors of the 
basis vectors so that four out of the five remaining components are real and 
non-negative. Thus we arrive at the canonical form 
\begin{equation}
|\psi\rangle = r_0e^{i\phi}|000\rangle +r_1|100\rangle +
                 r_2|010\rangle +r_3|001\rangle +r_4|111\rangle \ .
\label{acinB}
\end{equation}
\noindent Taking normalization into account only five free parameters remain. 
(A word of warning: The state $|\psi\rangle$ can be presented in this way if 
$|111\rangle$ is the closest separable state, but the converse does not hold.) We 
would like to know, for instance, what values the invariants 
have to take in order for a further reduction to four, three, or two, 
non-vanishing components to be possible, but for this purpose other choices of 
the canonical form are preferable \cite{AAJ01}.

We leave these matters here, and turn to the coarser classification based on stochastic 
LOCC (SLOCC), that is operations which locally transform the initial state into the
target state with a non-zero probability.
We will analyze the orbits arising when the group $G_L$
acts on given pure states. In contrast to the group $G_U$ of local unitaries, 
the group $G_L$ is not compact and the orbits may not be closed---one orbit can 
sit in the closure of another, and hence the set of orbits will have an intricate 
topology. 

The unitary invariants $I_1, \dots , I_5$ will not survive as invariants 
under the larger group $G_L$---unsurprisingly, since they were constructed 
using complex conjugation, and $G_L$ is the complexification of $G_U$. 
SLOCC operations can change even the norm $I_1$ of a state. We have one card left 
to play though, namely the hyperdeterminant ${\rm Det}_3(\Gamma )$. 
From its definition in terms of the tensor $\Gamma^{i_1i_2i_3}$ and the $\epsilon$-tensor, 
in Eq. (\ref{det33}), it is clear that it changes with a determinantal factor under 
SLOCC. Indeed 
\begin{equation}
 {\rm Det}_3(\Gamma ) = ({\rm det}L_1)^2 ({\rm det}L_2)^2 ({\rm det}L_3)^2 \;
{\rm Det}_3(\Gamma') \ . 
\label{detinvar}
\end{equation}

\noindent Hence the hyperdeterminant is a relative invariant, and a state for which it 
vanishes---such as the $W$ state---cannot be transformed into a state for which it is 
non-zero---such as the GHZ state.
\index{equivalence!SLOCC}
Pure states of three qubits can be entangled in two inequivalent ways. There is some 
additional fine structure left, and indeed a set of six independent covariants exist 
that together provide a full classification. Rather than listing them all Table 
\ref{tab:3qub} takes the easy way out, and uses the observation that the rank of 
the reduced density matrices, $r_A=r(\rho_A)=r({\rm Tr}_{BC}\rho_{ABC})$,
cannot be changed by an invertible SLOCC transformation, 
so if  $|\psi\rangle \equiv _{\rm SLOCC}  |\phi\rangle $
then all three local ranks have to be pairwise equal,
e.g. $r_A(|\psi\rangle)=r_A(|\phi\rangle)$. 
From the Table it is clear that the local ranks are enough to characterize a state as 
separable, or as containing bipartite entanglement only, but they cannot distinguish 
between the GHZ and $W$ states---which, as we know from our tour of the symmetric 
subspace in Section \ref{sec:permutat}, are SLOCC inequivalent. For this we need 
the hyperdeterminant. 

This division into SLOCC equivalence classes is complete. 
(This was first shown 
by D{\"u}r, Vidal and Cirac \cite{DVC00}. Actually, in a way, the classification had already been given by Gelfand et al. \cite{GKZ94} 
in their book. But they used very different words.)
The GHZ--class is dense in the set of all pure states while the W--class is of measure 
zero, and any state in the latter can be well approximated by a state in the former 
\cite{WRZ05}. This is an important point, 
and we will return to it in Section \ref{sec:polytopes}. 

\begin{table}[h]
\caption{SLOCC equivalence classes for three--qubit pure states $|\psi_{ABC}\rangle$:
  $r_A, r_B$ and $r_C$ denote ranks of single--partite reductions,
   while ${\rm Det}_3(\Gamma )$ is the hyperdeterminant
  of the $3$--tensor representing the state.}
 \smallskip
\hskip -0.5cm
{\renewcommand{\arraystretch}{1.21}
\begin{tabular}
[c]{cccccc} 
\hline \hline
Class &  $r_A$  & $r_B$ & $r_C$  & $|{\rm Det}_3(\Gamma )|$ & entanglement \\
  \hline \hline
                                separable  & $1$ & $1$ & $1$ &  $=0$ & none \\
$|\phi_A\rangle \otimes |\psi_{BC}\rangle $ & $1$ & $2$ & $2$ &  $=0$ & bipartite \\
$|\phi_B\rangle \otimes |\psi_{AC}\rangle $ & $2$ & $1$ & $2$ &  $=0$ & bipartite\\
$|\phi_C\rangle \otimes |\psi_{AB}\rangle $ & $2$ & $2$ & $1$ &  $=0$ & bipartite\\
                               $|W\rangle $ & $2$ & $2$ & $2$ &  $=0$ & triple bipartite\\
                             $|GHZ\rangle $ & $2$ & $2$ & $2$ &  $>0$ & global tripartite \\
\hline \hline
\end{tabular}
}
\label{tab:3qub}
\end{table}

How do these results generalize to larger systems? 
For any number of quNits the local unitary group $G_U=SU(N)^{\otimes K}$ is 
compact, it has closed orbits when acting on 
the complex projective space ${\mathbbm C}{\bf P}^{N^K-1}$, and Hilbert's 
theorem (and its extensions) guarantees that these orbits can be 
characterized by a finite set of invariants. However, it is already 
clear from our overview of the situation in the symmetric subspace 
(Section \ref{sec:permutat}) that the results would be so complicated 
that one may perhaps prefer not to know them. If instead we consider 
equivalence under SLOCC for modest numbers of qubits things do look better.  
Polynomial invariants are known for four \cite{LT03,Le06}
and five qubits \cite{LT06}, although the results for the latter are 
partial only. In the $4$-qubit Hilbert space there exists infinitely 
many SLOCC orbits, and their classification is not straightforward 
\cite{GW10,Ost10}. Still they can be organized into {\sl nine} families 
of genuinely four-partite entangled states, six of them depending on a 
continuous parameter \cite{VDDV02, ChDj07}. We will give a slightly 
coarser classification of this case (into seven subcases) in Section 
\ref{sec:polytopes}. 

To go to higher numbers of qubits we definitely need a yet coarser classification 
of entangled states. This is where the hyperdeterminant comes into 
its own---although it will have occurred to the reader that we have not really 
\index{invariants!hyperdeterminant}
defined the hyperdeterminant, we simply gave an expression for it in the simplest case. 
To outline its geometrical meaning, let us return to the bipartite case, but for 
two $N$-level subsystems. The separable states are then described by the Segre 
embedding of ${\mathbbm C}{\bf P}^{N-1}\times {\mathbbm C}{\bf P}^{N-1}$ into 
${\mathbbm C}{\bf P}^{N^2-1}$. 
\index{embedding!Segre}
Any state can be described by a tensor $\Gamma^{ij}$, 
and generic entangled states can be characterized by the fact that its determinant 
is non-zero. For two quNits there are many intermediate cases for which the determinant 
vanishes even though the state is non-separable. In fact the rank of $\Gamma$ gives rise 
to an onion-like structure, with all the non-generic cases characterized by the single 
equation $\det{\Gamma} = 0$. To unravel its geometric meaning we have to recall the 
notion of projective duality from Section  4.1.   
The idea is that a hyperplane in ${\mathbbm C}{\bf P}^{N^2-1}$ can be regarded as a 
point in a dual copy of the same space. Now consider a hyperplane corresponding to 
the point $\bar{\Gamma}_{ij}$ in the dual space, and tangent to the separable Segre 
variety at the point $x^iy^j$. This hyperplane is defined by the equation 
\begin{equation} F = \bar{\Gamma}_{ij}x^iy^j = 0 \ , \end{equation} 

\noindent together with 
\begin{equation} \frac{\partial}{\partial x^i}F = \frac{\partial}{\partial y^i}F = 0 
\hspace{5mm} \Leftrightarrow \hspace{5mm} \bar{\Gamma}_{ij}y^j = x^j{\bar{\Gamma}}_{ji} = 0 
\ . \end{equation}

\noindent The second set of equations ensures that the hyperplane has `higher order contact' 
with the Segre variety, as befits a tangent hyperplane. But the condition that these 
equations do have a solution is precisely that $\det{\bar{\Gamma}} = 0$. This gives a 
geometrical interpretation of this very equation. It only remains to find the generalization 
to the set of tangent planes of the generalized Segre variety ${\mathbbm C}{\bf P}^{N-1} \times 
{\mathbbm C}{\bf P}^{N-1} \times \dots \times {\mathbbm C}{\bf P}^{N-1}$. 

This is not an easy task. For three qubits it turns out to be precisely 
the condition that the hyperdeterminant, as given above, vanishes. A similar polynomial 
equation (in the components of the tensor that defines the state) exists for any number of 
qubits. The polynomial is called the hyperdeterminant, and its degree is known. Unfortunately 
the latter raises quickly. For the case of four qubits the  
degree is $24$ \cite{GKZ94}. It 
is indeed useful to quantify four--qubit entanglement \cite{LT03,Mi03},
and it can be represented as a function of
four other invariants of a smaller degree \cite{DO09,VES11},
including the determinants of two--qubit reduced states. For larger number of qubits---and 
for quNits---it is hard to be explicit about this, but at least it is comforting to know 
that such a classification exists.

\section{Monogamy relations and global multipartite entanglement}
\label{sec:monogamy}

Any system consisting of three subsystems can be split into 
two parts in three different ways. Furthermore
one can consider any two parties and investigate their entanglement.
How is the entanglement
between a single party $A$ and the composite system $BC$, written $A|BC$, related to 
the pairwise entanglement $A|B$ and $A|C$?

Coffman, Kundu and Wootters analysed this question \cite{CKW00}
using tangle as a measure of entanglement.
Recall that this quantity equals the square of the concurrence, $\tau(\rho)=C^2(\rho)$,
and is known analytically for any mixed state of two qubits \cite{Wo98}. 
See Eq. (16.103). 
They established the {\it monogamy relation} 
\begin{equation}
\tau_{A|BC} \ge \tau_{A|B}  + \tau_{A|C} \ .
\label{monogamy}
\end{equation}
Here $\tau_{A|B}$ denotes the tangle
of the two--qubit reduced state, 
$\rho_{AB}={\rm Tr}_C \rho_{ABC}$,
while $\tau_{A|BC}$ represents the tangle 
between part $A$ and the composite system $BC$.
Although the state of subsystems $BC$ lives in four dimensions,
its rank is not larger than two, 
as it is obtained by the reduction of the pure state 
$\rho_{ABC}=|\psi_{ABC}\rangle \langle \psi_{ABC}|$.
This observation allows one to 
describe entanglement along the partition 
$A|BC$ with the two--qubit tangle (16.103) 
and to establish inequality (\ref{monogamy}).

Even though subsystem $A$ can be simultaneously 
entangled with the remaining subsystems $B$ and $C$,  
the sum of these two entanglements
cannot exceed the entanglement between $A$ and $BC$.
This implies, for instance, that if $\tau_{A|BC} =\tau_{A|B}=1$
then $ \tau_{A|C}=0$.
Hence if the qubit $A$ is maximally entangled with $B$,
then it cannot be entangled with $C$.

To characterize entanglement in the three--qubit systems
it is natural to consider the tangle averaged over all possible splittings.
The average tangle of a pure state $\rho_{ABC}=|\psi_{ABC}\rangle\langle \psi_{ABC} |$ 
with respect to $1+2$ splittings reads
\begin{equation}
\tau_1\bigl( |\psi_{ABC}\rangle \bigr) \equiv \frac{1}{3} \bigl (\tau_{A|BC} + \tau_{B|AC} + \tau_{C|AB}\bigr) \ .
\label{tau1}
\end{equation}
Another related quantity describes average entanglement contained in two-partite reductions, 
\begin{equation}
\tau_2\bigl( |\psi_{ABC}\rangle \bigr) \equiv \frac{1}{3} \bigl (\tau_{A|B} + \tau_{B|C} + \tau_{C|A}\bigr) 
\ .
\label{tau2}
\end{equation}

\noindent By construction, the quantities $\tau_1, \tau_2$ are non-negative
and for any pure state can be computed analytically 
with help of the Wootters formula  (16.103). 
As entanglement is monotone with respect to partial trace,
 $\tau_{A|BC} \ge \tau_{A|B}$
(see Section 16.8) 
the relation 
$ \tau_1\bigl( |\psi_{ABC}\rangle \bigr) \ge \tau_2\bigl( |\psi_{ABC}\rangle \bigr)$
holds for every pure state. 

It is easy to check that $\tau_1$ achieves its maximum for the GHZ state,
\begin{equation}
0 \le \tau_1\bigl( |\psi_{ABC}\rangle \bigr)  \le 
\tau_1\bigl( |GHZ \rangle \bigr) =1 \ .
\label{tau1bound}
\end{equation}

\begin{figure} [h]
   \begin{center} \
 \includegraphics[width=9.5cm,angle=0]{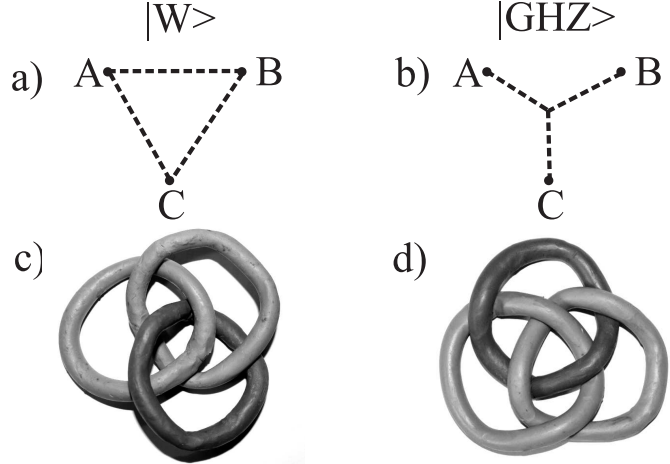} 
\caption{Schematic comparison between the distinguished 3--qubit 
pure states with the help of strings and knots:
if one subsystem is traced away from a system
in the $W$ state (panels a and c), the other two subsystems remain entangled.
This is not the case for the GHZ state,
thus it can be represented by three objects joined with a single thread
or three {\it Borromean rings},
which enjoy an analogous property. See panels b) and d).}
\label{fig:ghz_w}
\end{center}
 \end{figure}

\noindent As discussed in Problem 17.1 
after tracing out any single part of $|GHZ\rangle$,
the remaining two subsystems become separable, so 
$\tau_2\bigl( |GHZ \rangle \bigr) =0$. See a cartoon sketch in Figure \ref{fig:ghz_w}.
The latter quantity,
characterizing the mean  bipartite entanglement,
 is maximized by the $W$ state \cite{DVC00},
\begin{equation}
0 \le \tau_2\bigl( |\psi_{ABC}\rangle \bigr)  \le 
\tau_2\bigl( |W \rangle \bigr) =4/9.
\label{tau2bound}
\end{equation}
Thus taking into account the quantity $\tau_1$ the GHZ state
is `more entangled' than the $W$ state, 
while the opposite is true if we look at the measure $\tau_2$. 
The fact that the maximum in (\ref{tau2bound}) is smaller than unity
is another feature of the monogamy of entanglement:
an increase of the entanglement between $A$ and $B$
measured by tangle
has to be compensated by a corresponding decrease of 
the entanglement between $A$ and $C$ or $B$ and $C$.

%

The last, and the most important, measure, 
which characterizes the {\it global entanglement},  
is the {\it $3$--tangle} \cite{CKW00},
\begin{equation}
\tau_3\bigl( |\psi_{ABC}\rangle \bigr) \equiv  
\tau_{A|BC} - \tau_{A|B} - \tau_{A|C} .
\label{tau3}
\end{equation}
This quantity, invariant with respect to permutation
of the parties, is also called {\it residual entanglement},
as it marks the fraction of entanglement which cannot be 
described by any two-body measures.

For any three qubit pure state $|\psi_{ABC}\rangle$ 
written in terms of the components $G^{ijk}$
one can use the Wootters formula (16.103) 
for tangle and 
find an analytical expression for the $3$-tangle. It turns out \cite{CKW00,DVC00}
that the three--tangle is proportional to the modulus of the hyperdeterminant,
\index{invariants!hyperdeterminant} 
\begin{equation}
\tau_3\bigl( |\psi_{ABC}\rangle \bigr) = 4 |{\rm Det}_3(G)| ,
\label{tau3bis}
\end{equation}
which already appeared in (\ref{det33}).
Thus  $\tau_3$ is indeed invariant with respect to local unitary operations,
as necessary for an entanglement measure
and forms also an entanglement monotone \cite{DVC00}. 
\index{entanglement!three--tangle}
Furthermore, $\tau_3$ is a relative invariant under the action of 
the group $G_L=GL(2,{\mathbb C})^{\otimes 3}$, so it 
distinguishes the class of $W$ states
and the GHZ states with respect to SLOCC transformations.
Due to the monogamy relation (\ref{monogamy})
the 3--tangle is non-negative, and it is equal to zero
for any state separable under any cut.
Its maximum is  achieved for the GHZ state,
for which  $\tau_{A|BC}=1$ and $\tau_{A|B}=\tau_{A|C}=0$,
so that $\tau_3\bigl( |GHZ \rangle \bigr) =1$.
With respect to the Fubini--Study measure on the space of 
pure states of a three--qubit system it is possible to obtain \cite{KNM02}
the average value $\langle \tau_3 \rangle_{\psi} =  1/3$.

The monogamy relation (\ref{monogamy}),
originally established for three qubits \cite{CKW00},
was later generalized for several qubits by Osborne and Verstraete \cite{OV06},
while Eltschka and Siewert \cite{ES15}
derived monogamy equalities---exact relations 
between different kinds of entanglement  satisfied by pure states 
of a system consisting of an arbitrary number of qubits.

A generalization to quNits, with $N > 2$, is even more tricky.
Monogamy relation based on tangle does not hold
for several subsystems with three or more levels each \cite{Ou07}.
Such relations can be formulated for 
negativity extended to mixed states by convex roof \cite{KDS09}
and squared entanglement of formation \cite{OCF14}.
Furthermore, for systems of an arbitrary dimension
general monogamy relations hold for the squashed entanglement \cite{KGS12},
as this measure is known to be additive.

Observe that the tangle of a pure state of two qubits can be written as 
$\tau(\psi) = |\langle \psi |   \sigma_y \otimes \sigma_y |\psi^*\rangle|^2$,
where $|\psi^*\rangle$ denotes the state after 
complex conjugation in the computational basis.
In a similar way for a four--qubit pure state
one can construct the four--tangle
$\tau_4(\psi) = |\langle \psi |   \sigma_y^{\otimes 4} |\psi^*\rangle|^2 $.
This quantity, introduced by Wong and Christiansen \cite{WC01},
can be interpreted as $4$-party residual entanglement
that cannot be shared between two--qubit bipartite cuts \cite{GW10}.
Four--tangle is invariant with respect to permutations and 
extended for mixed states by convex roof forms an entanglement monotone \cite{WC01}.

\section{Local spectra and the momentum map}
\label{sec:polytopes}

The story of this section begins with an 
after-dinner-speech by C. A. Coulson, who observed that the most interesting 
properties of many-electron systems tend to concern observables 
connecting at most two parties. To calculate them, only the reduced two-party 
density matrices are needed. 
(Coulson's question (1960) 
\cite{Co60} inspired important work by A. J. Coleman \cite{Co63,CY00}. Their 
version of the problem concerned fermionic constituents. We consider 
distinguishable subsystems only.)

The question he raised was: what conditions on a 
two--party density matrix ensure that it can have arisen as a partial trace over 
$K-2$ subsystems of a pure $K$-partite state? This is a version of the 
{\it quantum marginal problem}. In the analogous classical problem 
one asks for the conditions on a pair distribution $p_{12}^{ij}$ (say)  
ensuring that it can arise as a marginal distribution from the joint distribution 
$p_{123}^{ijk}$. This is a question about the projection of a simplex in 
${\mathbbm R}^{N^3}$ to a convex body in ${\mathbbm R}^{N^2}\oplus 
{\mathbbm R}^{N^2}\oplus 
{\mathbbm R}^{N^2}$. After finding the pure points of the projected body 
one has to calculate the facets of their convex hull. But this is a computationally 
demanding problem. Still the projection 
of a point is always a point, so the classical problem is trivial for pure states. 
Not so the quantum problem. Already when one starts from a pure quantum state one needs 
sophisticated tools from algebraic geometry, and from invariant theory, in order to set 
up an algorithm for it. Finding the restrictions on the resulting two-party density 
matrices is really hard. 
The problem was reduced to that of describing 
momentum polytopes (see below) by Klyachko \cite{Kl04} 
and by Daftuar and Hayden \cite{DH05}, in both cases relying on earlier work 
by Klyachko \cite{Kly98}. 

For bipartite pure states the 
story is simple. There are two reduced density matrices $\rho_A={\rm Tr}_B \rho$
and $\rho_B={\rm Tr}_A \rho$, and their spectra are the same (up to extra zeros, 
if the two subsystems differ in dimensionality). That spectrum can be anything 
however, which is dramatically different from the classical case where the 
subsystems are in pure states whenever the composite system is. To figure out what 
happens for 3-qubit systems we choose a partition, say $A|BC$, and perform a 
Schmidt decomposition. Thus 
\begin{equation} |\psi_{ABC}\rangle = \sqrt{\lambda_A}\sum_{i,j}a^{ij}|0_Ai_Bj_C
\rangle + \sqrt{1-\lambda_A}\sum_{i,j}b^{ij}|1_Ai_Bj_C\rangle \ . \end{equation}

\noindent By construction 
\begin{equation} \sum_{i,j}a^{ij}\bar{a}_{ij} = \sum_{i,j}b^{ij}\bar{b}_{ij} = 1 \ , 
\hspace{8mm} \sum_{i,j}a^{ij}\bar{b}_{ij} = 0 \ . \label{aborto} \end{equation}
  
\noindent As always we are free to choose the local bases, and this we do in 
such a way that $|0_A\rangle$ is the eigenstate of $\rho_A$ with the smallest 
eigenvalue $\lambda_A$, and similarly for {\sl all three} subsystems. We can then 
read off that 
\begin{equation} \lambda_B = \lambda_A\sum_ja^{0j}\bar{a}_{0j} + 
(1-\lambda_A)\sum_jb^{0j}\bar{b}_{0j} \ , \end{equation}

\noindent and similarly for $\lambda_C$. From this we find the inequalities 
\begin{eqnarray} \lambda_B + \lambda_C = \lambda_A(a^{0j}\bar{a}_{0j} + 
a^{i0}\bar{a}_{i0}) + (1-\lambda_A)(b^{0j}\bar{b}_{0j} + 
b^{i0}\bar{b}_{i0}) \geq \nonumber \\ \nonumber \\ 
\geq \lambda_A\left( \sum_{i,j}a^{ij}\bar{a}_{ij} - |a_{11}|^2\right) + 
(1-\lambda_A)\left( \sum_{i,j}b^{ij}\bar{b}_{ij} - |b_{11}|^2\right) \geq \\ \nonumber \\
\geq \lambda_A\left( 2-|a_{11}|^2 - |b_{11}|^2\right) \ , \hspace{30mm} \nonumber \end{eqnarray}

\noindent where we used $\lambda_A \leq 1-\lambda_A$ and Eqs. (\ref{aborto}) 
in the last step. However, $a_{11}$ and $b_{11}$ are the corresponding components 
of two orthonormal vectors, and hence they obey $|a_{11}|^2 + |b_{11}|^2 \leq 1$. 
(This is easy to prove using the Cauchy-Schwarz inequality, perhaps avoiding the 
composite indices for clarity.) Thus, repeating the exercise for all three 
partitions, we arrive at the triangle inequalities obeyed by the smallest 
eigenvalues of the three reduced density matrices, namely  
\begin{equation}
\lambda_A  \le  \lambda_B + \lambda_C \ ; \ \ \ 
\lambda_B  \le  \lambda_A + \lambda_C \ ; \ \ \ 
\lambda_C  \le  \lambda_A + \lambda_B \ . 
\label{triangle}
\end{equation}

\noindent Hence there are definite restrictions on the local spectra in the 
3-qubit case. 

There are no other restrictions. To see this, consider the state 
\begin{equation} |\psi\rangle = a|001\rangle + b|010\rangle + c|100\rangle 
+ d|111\rangle \ , \label{family3} \end{equation}

\noindent where all the components are real. 
Straightforward calculation verifies that 
\begin{equation} \left\{ \begin{array}{l} \lambda_A = a^2 + b^2 \\ 
\lambda_B = a^2+c^2 \\ \lambda_C = b^2+c^2 \end{array} \right. \hspace{5mm} 
\Rightarrow \hspace{5mm} \left\{ \begin{array}{ll} 2a^2 = \lambda_A + \lambda_B 
- \lambda_C \\ 2b^2 = \lambda_A + \lambda_C - \lambda_B \\ 2c^2 = 
\lambda_B + \lambda_C - \lambda_A & . \end{array} \right. \end{equation}

\noindent By choosing $a,b,c$ we can realize any triple $(\lambda_A, 
\lambda_B,\lambda_C)$ obeying the triangle inequalities. 

At the expense of some notational effort the argument can 
be repeated for $K$-qubit systems, resulting in {\it polygon inequalities} of the form  
\begin{equation} \lambda_k \leq \lambda_1 + \dots + \lambda_{k-1} + \lambda_{k+1} 
+ \dots + \lambda_K \ . \label{polygonineq} \end{equation}

\noindent Here $\lambda_k$ denotes the smallest eigenvalue of the reduced density matrix 
for the $k$-th subsystem. Again the inequalities are sharp: no further 
restrictions occur. 
Full details can be 
found in the paper by Higuchi, Sudbery, and Szulc \cite{HSS03}, from which the 
whole argument is taken. Bravyi \cite{Bra04} provides some further results. 

\begin{figure} [htbp]
   \begin{center} \
 \includegraphics[width=13.1cm,angle=0]{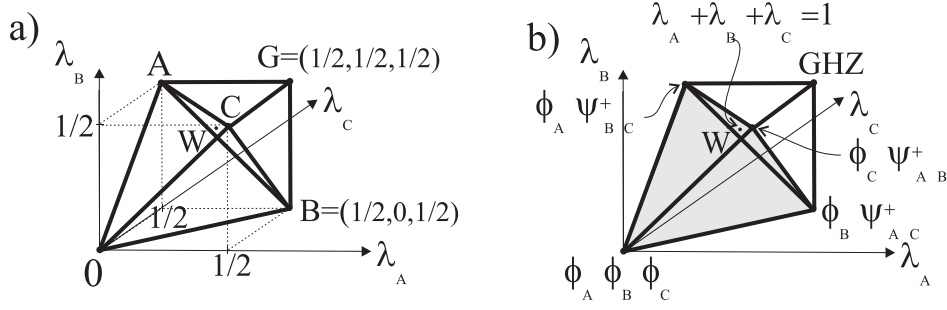}
\caption{Entanglement polytope for three qubits:
a) it is formed by three smaller eigenvalues 
of the  single--party reduced states,
 satisfying the triangle inequalities (\ref{triangle}); 
b) Corners $O$ and $G$ represent fully 
 separable and GHZ states, respectively.
  Three edges of the polytope connected to $O$ denote states
 with bipartite entanglement only. States equivalent with 
respect to SLOCC to $W$ belong to the shaded pyramid. 
Glueing eight such bipyramids together around the GHZ point 
we obtain a stellated octahedron or 
`stella octangula', which is not convex. Compare Figure 11.3.
}
\label{fig:kirw}
\end{center}
 \end{figure}

Together with the inequalities $0 \leq \lambda_k \leq 1/2$ the polygon inequalities 
define a convex polytope known as the {\it entanglement polytope}, or sometimes as 
the {\it Kirwan polytope} (for a reason we will come to). It is the convex hull of 
$2^K - K$ extreme points which are easily found since the whole polytope is 
inscribed in a hypercube with corners whose $K$ coordinates equal either $0$ or $1/2$. 
Let us refer to the corner $(0,0, \dots , 0)$ as the separable corner, since separable 
states end up there. The inequalities (\ref{polygonineq}) imply that all the $K$ corners 
at Hamming distance 1 from the separable corner are missing, whereas all the other corners 
of the cube are there. On the long diagonal connecting the separable corner to the 
GHZ corner $(1/2, 1/2, \dots , 1/2)$ --- which is where the image of the GHZ state is 
to be found---we find the images of states in the symmetric subspace. 
The case $K = 3$ is an easily visualized bipyramid with 5 corners and 6 faces (Figure 
\ref{fig:kirw}). When $K = 4$ there are 12 corners and 12 facets; the latter are of three 
different kinds and include 4 copies of the $K = 3$ bipyramid. When $K = 5$ there are 
27 corners and 15 facets, and so it goes on. 

The story becomes much more interesting once we ask where differently entangled states 
land in the polytope. For three qubits the story is simple. The $W$ state lands at 
$(1/3,1/3,1/3)$, and for all states that are SLOCC equivalent to the $W$ state one 
finds 
\begin{equation} \lambda_A + \lambda_B + \lambda_C \leq 1 \ . \end{equation}

\noindent In fact the image of this class of states forms a polytope of its own, 
making up the lower pyramid in Figure \ref{fig:kirw}. In this particular case 
elementary arguments suffice to prove it \cite{HZG04}. One can also show that 
states with bi-partite entanglement only form the three edges emanating from the 
separable corner. These edges are polytopes of their own, which we can call 
${\cal P}_{A|BC}$, ${\cal P}_{B|AC}$, and ${\cal P}_{C|AB}$. The image of the 
separable states is a single point ${\cal P}_{\rm sep}$. Images of the generic 
states (SLOCC equivalent to the GHZ state) form an open set whose closure is 
the entire polytope. In this way we have a hierarchy of polytopes, 
${\cal P}_{\rm sep} \subseteq {\cal P}_{A|BC} \subseteq {\cal P}_W \subseteq 
{\cal P}_{GHZ}$. A polytope in the hierarchy is a subpolytope of another if it 
corresponds to an orbit that lies in the closure of the other orbit, or in 
physical terms if a state from the former can be approximated by a state from 
the latter with arbitrary precision. 
   
We have learned that the local spectra carry some information about the kind of 
entanglement: if $\lambda_A + \lambda_B + \lambda_C > 1$ the state investigated 
is of the GHZ type. (Perhaps we should note though that the spectra of the set 
of all mixed states will fill the entire cube, so if we want to use the local 
spectra as witnesses of GHZ--type entanglement we need a guarantee that the 
states we measure are close to pure.) For a state of the GHZ class there is a two 
parameter family of orbits under local unitaries ending up at the same point 
in the entanglement polytope, while a pure state in 
the closure of the $W$ class is determined uniquely up to local unitaries by 
its local spectra \cite{SWK13}. Generically this ambiguity can be resolved by 
considering also the spectra of all bipartite reduced density matrices, 
but for Schmidt decomposable states some ambiguity remains \cite{LPW02}: For 
states of the form $a|000\rangle + b|111\rangle$ the spectrum of every reduced 
density matrix agrees with that obtained from the mixed state 
$|a|^2|000\rangle \langle 000| + |b|^2|111\rangle \langle 111|$.

The four dimensional entanglement polytope characterizing the four--qubit system 
was analysed in detail by Walter et al. \cite{WDGC13}, and by Sawicki et al. 
\cite{SOK14}. Their arguments rely on invariant theory and symplectic geometry, 
and are no longer elementary. (The first step in one of these papers is 
to calculate the 170 independent covariants of the four-qubit system \cite{BLT03} 
for the nine different families found by Verstraete et al. 
\cite{VDDV02}.) The results are easy to describe though, and are summarized in 
Table \ref{tab:gross}. There are six kinds of four-dimensional subpolytopes, 
partially ordered by the inclusion relations 
${\cal P}_{5} \subseteq {\cal P}_{3} \subseteq {\cal P}_2 \subseteq 
{\cal P}_{1}$ and ${\cal P}_{5} \subseteq {\cal P}_{4} \subseteq {\cal P}_6$. 
In addition to those we find lower dimensional subpolytopes describing 
states without genuine four-partite entanglement. This includes four facets 
(called $F_2$ in the Table) that are simply copies of the three-qubit 
polytope. Anyway the botany has become surveyable.  

\begin{table}[h]
\caption{Entanglement polytope for four qubits. The vertex $1111$ corresponds to 
the spectrum $(1/2,1/2,1/2,1/2)$, 
the 3-partite entangled vertices 
are $111\bar{1}$ etc., the 2-partite entangled vertices are 
$11\bar{1}\bar{1}$ etc., and the separable vertex 
is $\bar{1}\bar{1}\bar{1}\bar{1}$. The extra vertex $V$ in one of the subpolytopes 
corresponds to the spectrum $(1/4,1/2,1/2,1/2)$. There are three kinds of 
facets ($F_i$) 
and six kinds of full-dimensional subpolytopes (${\cal P}_i$). The number of permutation 
equivalent copies of each is given at the bottom.}
 \smallskip
\hskip -0.5cm
{\renewcommand{\arraystretch}{1.1}
\begin{tabular}
[c]{|c|ccc|cccccc|} 
\hline
Vertex & $F_1$  & $F_2$ & $F_3$ & ${\cal P}_1$  & ${\cal P}_2$ 
& ${\cal P}_3$  & ${\cal P}_4$  & ${\cal P}_5$  & ${\cal P}_6$  \\
  \hline 
$1111$  & $\times$ &  &  &  & & & $\times$ & & $\times$ \\ \hline
$V$ & & & & $\times$ & & & & & \\ \hline 																
$111\bar{1}$ & $\times$ & $\times$ & & $\times$ & & & & & \\
$11\bar{1}1$ & $\times$ &  &  & $\times$ & $\times$ & & & & \\
$1\bar{1}11$ & $\times$ &  &  & $\times$ & $\times$ & $\times$ & & & $\times$ \\
$\bar{1}111$ &  &  &  & $\times$ & $\times$ & $\times$ & & & $\times$ \\ \hline
$11\bar{1}\bar{1}$ & $\times$ & $\times$ & $\times$ &  $\times$ & $\times$ & 
$\times$ & $\times$ & $\times$ & $\times$ \\
$1\bar{1}\bar{1}1$ & $\times$ & & $\times$ & $\times$ & $\times$ & $\times$ & 
$\times$ & $\times$ & $\times$ \\
$1\bar{1}1\bar{1}$ & $\times$ & $\times$ & & $\times$ & $\times$ &$ \times$ & 
$\times$ & $\times$ & $\times$ \\
$\bar{1}11\bar{1}$ & & $\times$ & & $\times$ & $\times$ & $\times$ & $\times$ 
& $\times$ & $\times$ \\
$\bar{1}1\bar{1}1$ & & & $\times$ & $\times$ & $\times$ & $\times$ & $\times$ 
& $\times$ & $\times$ \\
$\bar{1}\bar{1}11$ & & & & $\times$ & $\times$ & $\times$ & $\times$ & $\times$ 
& $\times$ \\ \hline 
$\bar{1}\bar{1}\bar{1}\bar{1}$ & & $\times$ & $\times$ & $\times$ & $\times$ & $\times$ 
& $\times$ & $\times$ & $\times$ \\ 
\hline
Perms & 4 & 4 & 4 & 4 & 4 & 6 & 1 & 1 & 6 \\ \hline
\end{tabular}
}
\label{tab:gross}
\end{table}    

Our story is not complete since we have said nothing about restrictions on 
the spectra when the partial trace is taken over two subsystems only. And 
an explicit generalization to quNits is not easy (although the case of three 
qutrits is manageable \cite{Hig03}). But for us there is a more burning question 
to discuss: Why are the results as simple as they are? Why are all the conditions 
we encountered given by linear inequalities on the spectra of the reduced 
density matrices?

In the case of pure states
of a system consisting of $K$ subsystems of size $N$ the entanglement 
polytope lives in ${\mathbbm R}^{K(N-1)}$, so we are dealing with a map 
\begin{equation} {\mathbbm C}{\bf P}^{N^K-1} \rightarrow {\mathbbm R}^{K(N-1)} 
\ . \end{equation}

\noindent At this point we have to take up a thread that 
we left dangling at the end of 
Section  13.5,    
 and explain the concepts of momentum maps 
and momentum polytopes. Their names suggest that they have something to do with 
momentum, and indeed as our motivating example (forcing us to take a detour 
through analytical mechanics) we choose angular momentum. On the phase space of 
a particle, with coordinates $q^i$ and $p_i$, there exist three functions $J_i = 
J_i(q,p)$ such that 
\begin{equation} \delta q^i = \{ q^i,\xi^kJ_k\} \ , \hspace{8mm} 
\delta p_i = \{ p_i,\xi^kJ_k\} \ , \label{lieact} \end{equation}
\begin{equation} \delta J_i = \{ J_i,\xi^kJ_k\} = \epsilon_{ik}^{\ \ \ j}
\xi^kJ_j \equiv \xi_i^{\ j}J_j \ . \label{liecons} \end{equation}

\noindent Here we should think of $\xi^i$ as a vector in the Lie algebra of the 
rotation group $SO(3)$, but we also rewrote it as a matrix $\xi_i^{\ j}$ using 
the natural matrix basis in the Lie algebra. 
For any given point $x$ in phase space the vector $J_i$ 
is a linear functional on the Lie algebra, meaning that $J_i\xi^i$ is a real number. 
Thus $J_i$ sits in a vector space which is dual to the Lie algebra. 
The rotation group acts on its own Lie algebra by means of conjugation, and it 
also acts on the dual vector space by means of what we will soon refer to as 
the coadjoint action. In 
general, any Lie group $G$ has a Lie algebra $\mathfrak{g}$. This is a vector space, 
and there exists a dual vector space $\mathfrak{g}^*$ such that for any element $\alpha$ 
in $g^*$ and any element $\xi$ in $\mathfrak{g}$ there exists a real number $\langle 
\alpha , \xi\rangle$. The Lie group acts on $\mathfrak{g}$ by conjugation, and this 
gives rise to an action on  $\mathfrak{g}^*$ as well. It is known as the 
{\it coadjoint action}, denoted $\mbox{Ad}^*_g$ and defined by
\begin{equation}  \langle {\rm Ad}^*_g\alpha , \xi \rangle = \langle \alpha , 
{\rm Ad}_{g^{-1}} \xi \rangle = \langle \alpha , g^{-1}\xi g\rangle \ . \end{equation}

\noindent For a compact group like $SO(3)$ the distinction between $\mathfrak{g}$ and 
$\mathfrak{g}^*$ is slight, and the bilinear form $\langle \ , \ \rangle$ is simply given 
by the trace of a product of two matrices. But we have reached a point of view from which 
Eq. (\ref{liecons}) is really quite remarkable.  On the one hand, the rotation 
group acts on the functions $J_i$ by means of canonical transformations of phase space. 
On the other hand, it acts on them through its adjoint action. And the equation says 
that these two actions are consistent with each other. This key observation motivates 
the definition of the {\it momentum map}, that we are now coming to.
(A standard reference for the momentum map is the book by Guillemin 
and Steinberg \cite{GS84}. Readers who want a gentle introduction may prefer to begin with the book by Springer \cite{Spr01}.)

In general, suppose that we have a manifold with a symplectic form $\Omega$ defined 
on it (as in Section 3.4) 
Let a Lie group $G$ act on it in the 
symplectic way. Thus to each group element $g$ there is a map $x \rightarrow \Phi_g(x)$ 
from the manifold to itself, preserving the symplectic form. Let there exist a 
map $\mu$ from the manifold to the vector space $\mathfrak{g}^*$ dual to the Lie 
algebra $\mathfrak{g}$ of $G$. The group acts there too through the coadjoint action. 
Then $\mu$ is a momentum map provided it is {\it equivariant}, 
\index{maps!momentum}
\begin{equation} \mu (\Phi_g(x)) = {\rm Ad}_g^* \mu (x) \ , \end{equation}

\noindent and provided that the vector field 
\begin{equation} \xi (x) = \frac{{\rm d}}{{\rm d}t}_{|t=0} \Phi_{e^{t\xi}}(x) \end{equation}

\noindent is the Hamiltonian vector field for the function 
$\mu_\xi = \langle \mu(x), \xi\rangle$. In our example the second condition is given 
(in different notation) by Eq. (\ref{lieact}), and the first by Eq. (\ref{liecons}). 
 
Complex projective space---the space of pure quantum states---is a symplectic manifold, 
and in fact we have come across an important example of a momentum map already, namely 
\begin{equation} (Z^0, Z^1, \dots , Z^n) \rightarrow 
\frac{1}{Z\cdot \bar{Z}}(|Z^0|^2, |Z^1|^2, \dots , |Z^n|^2) \ . \end{equation}

\noindent The map is from ${\mathbbm C}{\bf P}^{n}$ to ${\mathbbm R}^{n}$, 
and the image is a probability simplex. If we normalize the state, and write 
$Z^i = \sqrt{p_i}e^{i\nu_i}$, it looks even simpler. We know from Section 4.7
 that $p_i$ and $\nu_i$ are action-angle variables. The action 
variables $p_i$ generate the action of an abelian Lie group which is the direct 
product of $n$ copies of the circle group $U(1)$. Both the conditions for a momentum 
map are fulfilled. From the present point of view ${\mathbbm R}^n$ is the dual space 
of the Lie algebra of this group. Remarkably, the image of 
${\mathbbm C}{\bf P}^n$ is a convex subset there. So this is our first example 
of a momentum polytope.

A series of scintillating 
theorems generalize this simple observation.
(These theorems were motivated 
by the Schur-Horn theorem (Section 13.5)  
 and are due to Atiyah \cite{At82}, 
Guillemin and Sternberg \cite{GS84}, and Kirwan \cite{Kir84}. We remind the reader 
about Knutson's nice review \cite{Kn00}. Sawicki et al. \cite{SOK14} provide a good summary.)
 They concern a compact Lie group $G$ acting on a 
symplectic manifold $M$. The Lie group has a maximal abelian subgroup $T$ 
(`$T$' for torus), with a Lie algebra $\mathfrak{t}$ (the Cartan subalgebra, 
see Section 6.5). 
We then have

\smallskip
\noindent {\bf First convexity theorem.} {\sl If an abelian group $G$ admits 
a momentum map the image $\mu (M)$ of this momentum map is a convex polytope 
whose extreme points are the fixed points of the group action.}
\smallskip
 
\noindent For non-abelian groups this need not hold, but then we can divide 
$\mathfrak{t}$ into Weyl chambers (as in Section 8.5 ), 
\index{Weyl chamber}
and focus on the one containing vectors with positive entries in decreasing 
order. Call it 
$\mathfrak{t}_+$. Next we observe that because the momentum map is equivariant 
an orbit under $G$ in $M$ will be mapped into an orbit---called a {\it coadjoint 
orbit}---in the dual $\mathfrak{g}^*$ of the Lie algebra, 
\index{orbit!coadjoint}
and moreover each 
coadjoint orbit crosses the chosen Weyl chamber exactly once. 
(We saw this happen in Section 8.5.) 
We can then define a map $\Psi$ from $M$ to 
$\mathfrak{t}_+$ as the intersection of the image under the momentum map $\mu$ of 
an orbit in $M$ and the positive Weyl chamber $\mathfrak{t}_+$. 

\smallskip
\noindent {\bf Second convexity theorem.} {\sl Under the conditions stated the image 
$\Psi (M)$ is a convex polytope in $\mathfrak{t}_+$.}
\smallskip

\noindent The result holds for all symplectic manifolds, and the convex 
polytopes arising in this way are called {\it momentum polytopes}.
\index{polytope!momentum}
Note that the restriction to the positive Weyl chamber is important---if 
we drop it we can obtain images like the non-convex stella octangula mentioned 
in the caption of Figure \ref{fig:kirw}.

Let us glance back on our problem, to classify orbits under the local unitary 
group $G_U = SU(2)^{\otimes K}$ in the many-qubit Hilbert space. It has a maximal 
abelian subgroup and a Cartan subalgebra $\mathfrak{t}$. The local spectra define 
diagonal density matrices in the dual space $\mathfrak{t}^*$, so the setting is 
right to apply what we have learned about the momentum map. We need a little more 
though, since we are mainly interested in equivalence under the SLOCC group $G_L$, 
and this group is not compact. But it is the complexification of the compact 
group $G_U$. We can then rely on the following \cite{Ne84, Br87, GuSj06}:   

\smallskip
\noindent {\bf Third convexity theorem.} {\sl Let $G$ be the complexification 
of a compact group, and let it 
act on ${\mathbbm C}{\bf P}^n$. Let $G\cdot x$ be an orbit through a point $x$, 
and let $\overline{G\cdot x}$ be its closure. Let the map $\Psi$ be defined as above.  
Then

a) the set $\Psi (\overline{G\cdot x})$ is a convex polytope, 

b) there is an open dense set of points for which $\Psi (\overline{G\cdot x}) = 
\Psi ({\mathbbm C}{\bf P}^n)$,  

c) the number of such polytopes is finite.}
\smallskip

\noindent Concerning the proof we confine ourselves to the remark that invariant 
theory is present behind the scenes; the finiteness properties of the polytopes 
are related to the finitely generated ring of covariants. Anyway this 
proposition is the platform from which the results of this section have been 
derived, in the references we have cited \cite{SOK14, WDGC13}. From it 
the problem of classifying multipartite entangled states looks at least more 
manageable than before.

\section{AME states and error--correcting codes}
\label{sec:maxmult}

In the bipartite case a maximally entangled state is singled out by the fact that 
any other state can be reached from by it means of LOCC. Moreover, a natural way to 
quantify bipartite entanglement is to start with a large but fixed number of 
copies of a state, and ask how many maximally entangled states we can distill 
from them by means of LOCC. That is, maximally entangled states serve as a 
kind of gold standard. In the multipartite case, where there are many different 
kinds of entanglement, we must learn 
to be pragmatic if we want to talk about `maximal entanglement'. One possible way 
to proceed is to restrict our attention to the much simpler, bipartite entanglement 
that is certainly present. We can ask for the average bipartite entanglement of the 
individual subsystems, or perhaps for the amount of bipartite entanglement averaged 
over every possible bipartition of the system. Various choices of the measure of the 
bipartite entanglement can be made.  

A very pragmatic way to proceed is to average the linear entropy 
(from Section 2.7)  
over all the reduced one-partite density matrices. Since we 
assume that the global state of the $K$ qubits is pure this is an entanglement 
measure. Under the name of the {\it Mayer--Wallach measure} it is defined by 
\begin{equation} Q_1(|\psi \rangle ) = 2\Bigl{\langle} S_L \Bigr{\rangle} =
2\left( 1 - \frac{1}{K}\sum_{k=1}^K\mbox{Tr}\rho_k^2 \right) \ . \end{equation}

\noindent It is normalized so that $0 \leq Q_1 \leq 1$, with $Q_1 = 0$ if and only if 
the state is separable and $Q_1 = 1$ if and only if each qubit taken individually is in a 
maximally mixed state. 
(It was written in this form by Brennen \cite{Br03}. 
Mayer and Wallach wrote it differently \cite{MW02}.)

Now choose a subset $X$ of qubits, with $|X| = k \leq K - k$ members, and trace out the 
remaining $K - k$ qubits; the assumption that $2k \leq K$ will simplify some statements. 
There are 
$K!/k!(K-k)!$ such bipartitions altogether, and we can define the entanglement measures
\begin{equation} Q_k(|\psi \rangle ) = \frac{2^k}{2^k-1}\left( 1 - \frac{k!(K-k)!}
{K!} \sum_{|X| = k}\mbox{Tr}\rho_X^2\right) \ .  \end{equation}

\noindent Again $0 \leq Q_k \leq 1$, with $Q_k = 0$ if and only if the global state 
$|\psi\rangle$ is separable, and $Q_k = 1$ if and only if it happens that all the reduced density 
matrices are maximally mixed. 
(Scott introduced these measures with the cautionary 
remark that they `provide little intellectual gratification' \cite{Sc04}.)
For the $W$ and GHZ states we find
\begin{equation} Q_k(|W_K\rangle ) = \frac{2^{k+1}}{2^k - 1}\frac{(K-k)k}{K^2} \ , 
\hspace{8mm} Q_k(|GHZ_K \rangle ) = \frac{2^{k-1}}{2^k-1} \ . \end{equation}

\noindent 
For the $W_6$ state we note that $Q_1 < Q_3 < Q_2$, which may seem odd. 
Moreover one can find pairs of states such that $Q_k(|\psi \rangle ) > 
Q_k(|\phi \rangle )$ and $Q_{k'}(|\psi \rangle ) < Q_{k'}(|\phi \rangle )$, 
for some $k' \neq k$. So there is no obvious ordering of the states into 
more or less entangled. Rather the measures $Q_k$ capture different aspects 
of multipartite entanglement as $k$ is varied. Moreover, if one changes 
the linear entropy to, say, the von Neumann entropy in the definition of $Q_k$, 
one may change the ordering of the states also when $k$ is kept fixed.
 See Problem  17.7.       
Maximizing $Q_k$ over the set of 
all states is a difficult optimization problem, becoming computationally 
more expensive if we use the von Neumann entropy \cite{BPB+07, FFPP08}. In 
general, we do not obtain a convincing definition of `maximally entangled' in 
this way.     

States for which the upper bound $Q_k = 1$ is reached are called {\it $k$--uniform}. 
\index{states!k--uniform}
The GHZ state, 
like some other states we know (see Problem 17.3), 
is $1$--uniform for any number of qubits. 
A $k$--uniform state (with $k \geq 2$) is always $(k-1)$--uniform, since 
the partial trace of a maximally mixed state is maximally mixed. If all reduced 
density matrices that result when tracing over at least half of the subsystems 
is maximally mixed, the state is said to be {\it absolutely maximally entangled}, 
abbreviated {\it AME} \cite{HCLRL12}. 
\index{states!AME}
Every measure of bipartite entanglement will have to agree that AME states---if 
they exist---are maximally entangled. 
 
Let us move on from qubits to the general case of $K$ subsystems with $N$ 
levels each. In the product basis a pure state is described by
\begin{equation} |\psi \rangle = {\Gamma}^{i_1i_2 \dots i_K}|i_1i_2 \dots i_K 
\rangle \ , \label{tensor11} \end{equation}

\noindent where the indices run from 1 to $N$. If a bipartition 
is made the tensor can be described by two collective 
indices, $\mu$ running from 1 to $N^k$ and $\nu$ running from 1 to $N^{K-k}$. 
Again we assume that $k\leq K-k$, and write the state as
\begin{equation} |\psi \rangle = \Gamma^{\mu \nu}|\mu \nu \rangle \ . \end{equation}

\noindent Tracing out $K-k$ subsystems we obtain the reduced state
\begin{equation} \rho_X = (\Gamma \Gamma^\dagger )^{\mu}_{\ \mu^\prime} 
|\mu \rangle \langle \mu^\prime | \ . \end{equation}

\noindent This state is maximally entangled (Section 16.3) 
 if and 
only if the rectangular matrix $\sqrt{N^k}\Gamma^{\mu \nu}$ is a right unitary 
matrix (also known as an isometry), that is if and only if  
\begin{equation} N^K\Gamma \Gamma^\dagger = {\mathbbm 1}_{N^k} 
\ . \end{equation}

\noindent For a $k$--uniform state this has to be so for {\sl every} bipartition 
of the $K$ subsystems into $k + (K-k)$ subsystems. This is clearly putting a 
constraint on the tensor $\Gamma$ which becomes increasingly severe as $k$ grows. 
For an AME state, with an even number of subsystems, such tensors are known 
as {\it perfect} \cite{PYHP15}, while the corresponding reshaped matrices 
are known as {\it multiunitary} \cite{GALRZ15}.  

AME states do exist for 2, 3, 5, and 6 qubits. For 3 qubits they are the GHZ states. 
For 5 qubits an AME(5,2) state is
\begin{eqnarray} |\Phi_{2}^5\rangle = && |00000\rangle +  \\ 
+ && |11000\rangle + |01100\rangle + |00110\rangle + |00011\rangle + |10001\rangle - \nonumber \\
- && |11000\rangle - |01100\rangle - |00110\rangle - |00011\rangle - |10001\rangle - \nonumber \\
- && |11110\rangle - |01111\rangle - |10111\rangle - |11011\rangle - |11101\rangle \ . 
\nonumber \end{eqnarray} 
It was given in this form by Bennett et al. \cite{BVSW96}, 
and by Laflamme et al. \cite{LMPZ96} in a locally equivalent 
form with only eight terms in the superposition.

\noindent To see how this state arises, recall the multipartite Heisenberg group 
from Section 12.4,   
and in particular the notation used in Eq. (12.50).  
The cyclic properties of the state make it easy to see that it is left 
invariant by the operators 
\begin{eqnarray} G_1 = && XXZ{\mathbbm 1}Z \ , \ G_2 = ZXXZ{\mathbbm 1} \ , \ G_3 = 
{\mathbbm 1}ZXXZ \ , \ G_4 = Z{\mathbbm 1}ZXX 
\ , \nonumber \\ G_5 = && ZZZZZ \ .  \end{eqnarray}

\noindent Moreover these five group elements generate a maximal abelian subgroup 
of the Heisenberg group, having 32 elements. In the terminology of 
Section 12.6
this means that $|\Phi_2^5\rangle$ is a stabilizer state.    

For the 6 qubit AME state, see Problem 17.8.
AME states do not exist for K = 4 \cite{GBP98,HS00}, or 
for more than 6 \cite{HGS16,Sc04}, qubits. AME states built 
from four qutrits do exist. An example is \cite{He13} 
\begin{eqnarray}
|\Phi_3^4\rangle = 
&&|0000\rangle+|0112\rangle+|0221\rangle+\nonumber\\
+ &&|1011\rangle+|1120\rangle+|1202\rangle+\nonumber\\
+ &&|2022\rangle+|2101\rangle+|2210\rangle \ .
\label{AME43}
\end{eqnarray}

\noindent Glancing at this state we recognize the finite affine plane 
of order 3, constructed in Eq.(12.58).
The labels from the first 
two factors of the basis vectors are used to define the position in the 
array, and then the labels from the last two factors are labelling the 
remaining two sets of parallel lines.
There are a number of  combinatorial ideas one can use to construct 
highly entangled states \cite{GALRZ15,GZ14}.
 It is not hard to show (Problem 17.9) 
that this is a stabilizer state. 

The same construction, using an affine plane of order $N = 2$, gives the 
Svetlichny state discussed in Problem 17.5.
For $N > 2$ the combinatorics of the affine plane ensures that we obtain a 
projector onto an $N^2$ dimensional subspace whenever we trace out two of 
the subsystems, which means that we obtain 2--uniform states of $N+1$ 
quNits---provided an affine plane of order $N$ exists, as it will if $N$ 
is a prime number, or a power of a prime number. Using results from 
classical coding theory one can show that AME states exist for any number 
of subsystems provided that the dimension of the subsystems is high enough 
\cite{HC13}. For qubits, we can still 
ask for the highest value of $k$ such that $k$--uniform states exist, and 
some asymptotic bounds are known for this. It is also known that if we 
restrict our states to the symmetric subspace, the upper bound for qubits is 
$k=1$ \cite{AC13}. So we cannot ask what AME states look like in the stellar 
representation, because it contains none (beyond GHZ$_3$). 

We now turn to this section's other strand: {\it error--correcting codes}. 
We begin classically, with the problem of sending four classical bits 
through a noisy channel. (Perhaps the message is sent by a space probe far out 
in the solar system.) Assume that there is a certain probability $p$ for any given bit 
to be flipped, and that these errors happen independently. In technical 
language, this is a {\it binary symmetric channel}. It may not be an accurate 
model of the noise. For instance, real noise has a tendency to come in bursts. 
But we adopt this model, and we want to be able to detect and correct the 
resulting errors. A simple--minded solution is to 
repeat the message thrice. Thus, instead of sending the message $0101$ we 
send the message $010101010101$. We can then correct any single error by 
means of a majority vote. But since the length of the message has increased, 
the probability that two errors occur has increased too. The trade-off is 
studied in Problem 17.10.  

It pays to adopt a geometric point of view. We regard each bitstring of length 
$n$ as a vector in the discrete vector space ${\mathbb Z}_2^n$ over the finite 
field ${\mathbb Z}_2$ (the integers modulo 2; 
see Section 12.2). 
Our repetition code is a subspace of dimension four in ${\mathbb Z}_2^{12}$. 
This information is summarized by the {\it generator matrix} of the code, in 
this case
\begin{equation} G_{(12,4)} = \left[ \begin{array}{cccc|cccccccc} 
1 & 0 & 0 & 0 & 1 & 0 & 0 & 0 & 1 & 0 & 0 & 0  \\ 
0 & 1 & 0 & 0 & 0 & 1 & 0 & 0 & 0 & 1 & 0 & 0 \\ 
0 & 0 & 1 & 0 & 0 & 0 & 1 & 0 & 0 & 0 & 1 & 0 \\ 
0 & 0 & 0 & 1 & 0 & 0 & 0 & 1 & 0 & 0 & 0 & 1  
\end{array} \right] \ . \label{repetition} \end{equation}

\noindent The four row vectors form a basis for the code, i.e. for a linear subspace 
containing altogether $2^4$ vectors. 
Their first four entries carry the actual information we want to send. 
The reason why this code enables us to correct single errors can now be expressed 
geometrically. Define the {\it weight} of a vector as the number of its non-zero 
components. One can convince oneself that the minimal weight of any non-zero vector 
in this code is 3. Therefore the minimal Hamming distance between any pair of vectors in 
the code is 3, 
because by definition the Hamming distance between two vectors ${\bf u}$ and ${\bf v}$ 
is the weight of the vector ${\bf u} - {\bf v}$, which is 
necessarily a vector in the code since the latter is a linear subspace. Thus, 
what the embedding of ${\mathbb Z}_2^4$ into ${\mathbb Z}_2^{12}$ has achieved is 
to ensure that the $2^4$ code words are well separated in terms of Hamming distance. 
Suppose a single error occurs 
during transmission, somewhere in this string. This means that we receive a vector 
at Hamming distance 1 from the vector ${\bf u}$. But the Hamming distance between two 
code words is never smaller than 3. If we surround each code vector with a `ball' of 
vectors at Hamming distance $\leq 1$ from the centre, we find that the vector received 
lies in one and only one such ball, and therefore we know with certainty which code 
word was being sent. This is why we can correct the error. 

In general, a linear {\it classical} $[n,k,d]$ {\it code} 
is a linear subspace of dimension $k$ in a discrete vector space of dimension $n$, 
with the minimal Hamming distance between the vectors in the subspace equal to $d$. 
The repetition code is the code $[12,4,3]$. 
Can we improve on it? Indeed, improvements are easily found by 
consulting the literature.
Classical error--correcting codes have been 
much studied since the time of Shannon's breakthrough in communication theory 
\cite{Sh48}. A standard reference is the book by MacWilliams and Sloane \cite{MS77}. 
For a briefer account, see Pless \cite{Pl82}.
 
The generator matrix of the $[7,4,3]$ {\it Hamming code} is   
\begin{equation} G_{(7,4)} = \left[ \begin{array}{cccc|ccc} 1 & 0 & 0 & 0 & 0 & 1 & 1 \\ 
0 & 1 & 0 & 0 & 1 & 0 & 1 \\ 0 & 0 & 1 & 0 & 1 & 1 & 0 \\ 0 & 0 & 0 & 1 & 1 & 1 & 1 
\end{array} \right] \ . \label{hamming} \end{equation}

\noindent With this code, the bitstring $0101$ is encoded as the message $0101010$. 
One can convince oneself that the minimum Hamming distance between 
the vectors in the code is again 3 (and indeed that these balls fill all of 
${\mathbb Z}_2^7$), so that all single errors can be corrected by this code as well. 
Moreover, an elegant and efficient decoding procedure can be devised, so that we 
do not actually have to look through all the balls. To see how this works, write 
$G = [ {\mathbbm 1} | A]$, and introduce the {\it parity check matrix} 
$H = [- A^{\rm T} | {\mathbbm 1} ]$. By construction
\begin{equation} HG^{\rm T} = 0 \ . \end{equation}  

\noindent In fact $H{\bf u} = 0$ for every vector in the code subspace. Now suppose 
that an error occurs during transmission, so that we receive the vector ${\bf u} + 
{\bf e}$. Then 
\begin{equation} H({\bf u} + {\bf e}) = H{\bf e} \ . \end{equation}

\noindent It is remarkable, but easy to check, that if the error vector ${\bf e}$ is 
assumed to have weight 1 then it can be uniquely reconstructed from the vector 
$H{\bf e}$. The conclusion is that the Hamming code allows us to correct all errors 
of weight 1 without, in fact, ever inspecting the message itself. It is enough to 
inspect the {\it error syndrome} $H{\bf e}$.   
 
The Hamming code is clearly more efficient than the simple repetition code, which 
required us to send three times as many bits as the number we want to send. If more 
than one error occur during the transmission we can no longer correct it 
using the $[7,4,3]$ code, but there is a $[23,12,7]$ code, known as the {\it Golay} code, 
that can deal with three errors. And so on. In fact, for a binary symmetric channel 
Shannon proved, with probabilistic methods, that the probability of errors in the 
transmission can be made smaller than any preassigned $\epsilon$ if we choose the 
dimension of the code subspace and the dimension of the space in which it is embedded 
suitably (with an eye on the properties of the channel) \cite{Sh48}. 

Quantum error correction, of a string of qubits rather than bits, is necessarily a 
subtle affair, since it has to take place without gaining any information about 
the quantum state that is to be corrected.
The first quantum 
error--correcting codes took the community by surprise when they were presented by 
Shor \cite{Sho95} and Steane \cite{St96} in the mid 1990s. They were linked 
to the Heisenberg group soon after that \cite{CRSS97, Go97}, and the theory then 
grew quickly.

 Suppose that the initial state is $|\Psi\rangle$, belonging to some 
multipartite Hilbert space ${\cal H}_N^{\otimes K}$. In the course of 
time---for concreteness assume 
that the transmission is through time, in the memory of a quantum computer---the state 
will suffer decoherence due to the unavoidable interaction with the environment. 
There is nothing digital about this process. But we do assume that the state 
is subject to a CP map of the form (10.53). 
We then introduce a unitary 
operator basis (Section 12.1), 
and expand all the Kraus operators in this 
basis. For our purposes it is best to express this in the environmental representation. 
Thus what has happened is that 
\begin{equation} |\Psi\rangle |0\rangle_{\rm env} \rightarrow 
\sum_IE_I|\Psi \rangle |\psi_I \rangle_{\rm env}  \ . \end{equation}

\noindent The elements of the unitary operator basis are called $E_I$ here, since this 
basis is soon going to deserve its alternative name `error basis'. 
The environmental states $|\psi_I\rangle_{\rm env}$ are not assumed to be orthogonal 
or normalized, and at this stage the introduction of the error basis is a purely 
formal device. If it were true that 
\begin{equation} \langle \Psi|E_I^\dagger E_J|\Psi \rangle = \mbox{Tr}
|\Psi\rangle \langle \Psi |E_I^\dagger E_J = \delta_{IJ} \ , \label{QECC1} 
\end{equation}

\noindent then we would be in good shape. The elements of the error basis would then 
give rise to mutually exclusive alternatives, and a measurement could be devised so 
that the state collapses according to
\begin{equation} \sum_IE_I|\Psi \rangle |\psi_I \rangle_{\rm env} 
\rightarrow E_I|\Psi\rangle |\psi_I \rangle_{\rm env} \ . \end{equation}

\noindent Once we knew the outcome, we could apply the operator $E_I^\dagger$ 
to the state, and recover the (still unknown) state $|\Psi \rangle$. 

If we replace the pure state $|\Psi \rangle \langle \Psi |$ with the maximally mixed 
state, the second equality in Eq. (\ref{QECC1}) would hold, but the maximally mixed 
state is not worth correcting. However, this suggests that we should restrict both 
the noise (the set of allowed error operators $E_I$) and the state $|\Psi \rangle $ 
in such a way that the state behaves as the maximally mixed state as far as the 
allowed noise is concerned. Beginning with the noise, we assume that it can be 
expanded in terms of error operators of the form 
\begin{equation} E_I = E_1\otimes E_2 \otimes \dots \otimes E_K \ , \end{equation}

\noindent with at most $w$ of the operators on the right hand side not equal to 
the identity operator. This is an error operator of {\it weight} $w$. Physically 
we are assuming that the individual qubits are subject to independent noise, which 
is a reasonable assumption, and also that not too many of the qubits are affected 
at the same time. The operator $E_I^\dagger E_J$ will therefore have at most 
$2w$ factors not equal to the identity. Next, we assume that the state is a 
$2w$-uniform state. Coming back to Eq. (\ref{QECC1}), we can perform the trace over 
the $K-2w$ factors where the error operators contribute with just the identity, 
obtaining (for quNits) 
\begin{equation} \langle \Psi|E_I^\dagger E_J|\Psi \rangle = \mbox{Tr}
|\Psi\rangle \langle \Psi |E_I^\dagger E_J = \frac{1}{N^{2w}}\mbox{Tr}
E_I^{\prime \dagger} E^{\prime}_J \ , \label{QECC2} 
\end{equation}

\noindent where $E^{\prime}_I$ is the non-trivial part of $E_I$. It follows 
immediately from the definition of the error basis that we obtain the desired 
conclusion (\ref{QECC1}). Hence the $2w$-uniform state can be safely sent over 
this noisy channel, and $w$ errors can be corrected afterwards. 

This is only a beginning of a long 
story. In general, a {\it quantum error--correcting code} 
of distance $d$, denoted $[[K,M,d]]_N$, is an $N^M$ dimensional subspace of an 
$N^K$ dimensional Hilbert space, such that errors affecting only $(d-1)/2$ 
quNits can be corrected along the lines we have described. The nicest 
error basis of all, the Heisenberg group (Chapter 12),  
comes 
into its own when such codes are designed. We have seen that a $k$--uniform 
state of $K$ quNits is a $[[K,1,k+1]]_N$ quantum error--correcting code 
\cite{Sc04}, but for more information we refer elsewhere 
\cite{St98, Pre98, DMN13}. For us it is enough that we have pointed a moral: 
beautiful states have a tendency to be useful too.

\section{Entanglement in quantum spin systems}
\label{sec:largeny}

So far we have had, at the back of our minds, the picture of a bold experimentalist 
able to explore all of Hilbert space. In many--body physics that picture is to be 
abandoned. For concreteness, imagine a cubic lattice with `atoms' described 
as quNits at each lattice site. The Hamiltonian is such that only nearest 
neighbours interact. There are $K$ lattice sites altogether. We will always 
assume that $K$ is finite, but we may let $K$ approach Avogadro's number, 
say $K = 10^{23}$. Then the total Hilbert space under consideration has 
$N^{10^{23}}$ dimensions. This is enormous. 
No experimentalist is going to explore this in full detail. Indeed, Nature itself 
cannot have done so during the $10^{17}$ seconds that have passed since the creation 
of the universe. Still multipartite entanglement in such systems will be 
important.
Many good reviews exist \cite{AFOV08,TSGAAL15}, and a forthcoming book 
by Zeng, Chen, Zhou, and Wen \cite{ZCZW15} describes how quantum information meets 
quantum matter. In today's laboratories it is possible to design quantum many--body 
systems with desirable properties \cite{BDZ08}.
Indeed the tensor product structure of Hilbert space gives rise to an intricate 
geography of quantum state space, and we may hope to find the ground states of physically 
interesting systems in very special places. 

The kind of systems we will focus on are called {\it quantum spin systems}. 
Some notation: Subsets of lattice sites will be denoted $X, Y, \dots$, and the number 
of sites they contain will be denoted by $|X|, |Y|, \dots$. Such a number can be regarded 
as the volume of the subset. The complementary 
subsets are denoted by $\bar{X}, \bar{Y}, \dots$, meaning that the entire lattice 
is the union of $X$ and $\bar{X}$. The boundary of a region $X$ is denoted by 
$\partial X$. The number of edges in the lattice passing through the boundary is 
denoted $|\partial X|$, and is regarded as the {\sl area} of the boundary. Each lattice site 
is associated with a Hilbert space of finite dimension $N$, and the total Hilbert 
space is the tensor product of all them. The Hilbert space associated to a region $X$ 
is the tensor product of all the Hilbert spaces associated to sites therein.  

With our change of perspective in mind, let us first divide the lattice into two 
parts (somehow). Thus the Hilbert space is 
\begin{equation} {\cal H} = {\cal H}_X\otimes {\cal H}_{\bar{X}} = (\otimes_{x\in X} 
{\cal H}_x) \otimes (\otimes_{x \in \bar{X}}{\cal H}_x) \ . \end{equation}

\noindent Now consider the amount of bipartite entanglement that arises if 
the global state is a pure state $|\psi\rangle$ chosen at random according to the 
Fubini-Study measure. By tracing out the complementary set $\bar{X}$ we obtain the 
reduced state $\rho_X = \mbox{Tr}_{\bar{X}}|\psi \rangle \langle \psi|$ associated 
to the subsystem formed by the atoms in subset $X$. Assume that $|X|$ is smaller than  
$|\bar{X}|$. We can now apply the Page formula (15.73),  
 keeping 
in mind that the size of the system is $N^{|X|}$ and the size of the environment 
is $N^{|\bar{X}|}$. The result is 
 \begin{equation}
    \langle E(|\phi\rangle)_X \rangle =
\langle S(\rho_X) \rangle \approx |X| \ln N - \frac{1}{2} N^{|X|-|\bar X|}.
\label{entvolumeny}
 \end{equation}

\noindent The second term describes a negative correction 
which can be neglected if $|X| \ll |\bar X|$. The leading term on the other 
hand grows proportionally to the number $|X|$ of subsystems
in the region $X$. In this way we arrive at the following statement:

\medskip
\index{law!volume}
{\bf Volume law:} For a generic multipartite quantum state 
the entanglement between any subregion $X$ and a larger environment $\bar{X}$ 
scales as the {\sl volume} of the region, as measured by the
number $|X|$ of its subsystems.
\medskip

This is also how the thermodynamical entropy behaves for physical systems 
with short range interactions---except that the thermodynamical entropy may 
well vanish at zero temperature, while the entanglement entropy does not, 
so they are distinct. 
The hope is that the relevant low energy states in a many--body system 
are far from generic---and that they are more amenable to computer calculations 
than the generic states are. Indeed the number of parameters needed to describe a 
generic state grows exponentially with the number of atoms, and this poses 
a quite intractable problem in computer simulations.  

At this point a sideways glance on a seemingly very different part of physics 
is useful. In the classical theory of general relativity black holes are assigned an 
entropy proportional to their area. It is known as the {\it Bekenstein-Hawking} 
entropy, and as it stands it has no microscopic origin. In an attempt to provide 
one it was noticed that in certain situations the entanglement entropy also 
grows with area.
(This was first seen in 1983 by Sorkin \cite{Sor83}; 
the concrete calculation performed by him and his coworkers was in the context 
of a non-interacting quantum field theory \cite{BKLS86}. For a review of black hole 
thermodynamics see Wald's book \cite{Wald94}.)
This suggests that we should replace the volume law with:

\smallskip
\index{law!area}
{\bf Area law:} For physically important states of many--body quantum systems
described by local interactions the entanglement
between any subregion $X$ and a larger environment $\bar X$
scales as the {\sl area} of the region, as measured by the number of interactions 
across its boundary $\partial X$.
\smallskip

We will give a precise statement later, but first we want to understand 
what kinds of states that can give rise to it. 

The spatial dimension of the lattice matters here. We begin with a one 
dimensional lattice, an open chain with $K$ sites. To find a suitable representation 
of the states relevant to such a system we are going to apply the Schmidt decomposition (see Section 9.2) 
stepwise to a tensor carrying $K$ indices.
First we recall Eq. (9.14), 
which gives the singular value decomposition of an 
$N_1\times N_2$ matrix $C$ as $C = UDV$, where $D$ is a diagonal matrix with 
Schmidt rank $r \leq {\rm min}(N_1,N_2)$.  
The (in general) rectangular matrices $U$ and $V$ obey $U^\dagger U = VV^\dagger = 
{\mathbbm 1}_r$. We also define the matrix 
\begin{equation} \Lambda \equiv D^2 \ . \label{Vidb} \end{equation}

\noindent If we start from 
a normalized state then $\Lambda$ is simply the reduced density matrix in diagonal form. 

Now any $K$-index tensor with $N^K$ components can be viewed as an $N\times N^{K-1}$ 
matrix, and the singular value decomposition can be applied to it:
\begin{equation} \Gamma^{i_ii_2 \dots i_K} = \Gamma^{i_1|i_2 \dots i_K} = 
\sum_{a =1}^rA^{i_1}_{\ a}(D_1V_1)_{a}^{\ |i_2 i_3 \dots i_K} 
\ . \label{firststep} \end{equation}

\noindent The matrix of left eigenvectors $U_1$ was renamed $A$, because we focus 
on its column vectors $A^{i_1}$. It will be observed that 
\begin{equation} \sum_{i_1}(A^{i_1})^\dagger A^{i_1} = U_1^\dagger U_1 = {\mathbbm 1}_{r_1} 
\ , \end{equation}
\begin{equation} \sum_{i_1}A^{i_1}\Lambda (A^{i_1})^\dagger = \sum_{i_1}A^{i_1} D VV^\dagger 
D (A^{i_1})^\dagger = \Gamma^{i_1 i_2 \dots i_K}\bar{\Gamma}_{i_1i_2\dots i_K} = 1 
\ . \end{equation}

\noindent To get the final equality we had to assume that the state is normalized.  

In the next step the $r_1\times N^{K-1}$ matrix on the right hand side of Eq. (\ref{firststep}) 
is reshaped into an $r_1N\times N^{K-2}$ matrix, and the singular value decomposition again 
does its job: 
\begin{equation} \Gamma^{i_ii_2 \dots i_K} = \sum_{a_1 =1}^rA^{i_1}_{\ a_1}
(D_1V_1)_{\ a_1}^{i_2 \ \ |i_3 \dots i_K} = \sum_{a_1=1}^{r_1}\sum_{a_2=1}^{r_2}
A^{i_1}_{\ a_1}A^{i_2}_{a_1a_2}(D_2V_2)_{a_2}^{\ \ |i_3 \dots i_K} 
\ . \label{secondstep} \end{equation}

\noindent The matrix $U^{i_2}_{\ a_a|a_2}$ was renamed $A^{i_2}_{a_1a_2}$, and is 
regarded as a collection of $N$ matrices of size $r_1\times r_2$. This time we have  
\begin{equation} \sum_{i_2}(A^{i_2})^\dagger A^{i_2} = U_2^\dagger U_2 = {\mathbbm 1}_{r_2} 
\end{equation}
\begin{equation} \sum_{i_2}A^{i_2}\Lambda_2 (A^{i_2})^\dagger = \sum_{i_2}A^{i_2} D_2 
V_2V_2^\dagger D_2 (A^{i_2})^\dagger = 
\Lambda_1 \ . \end{equation}

\noindent Clearly this procedure can be iterated. 
We have arrived at
\index{theorem!Vidal}

\smallskip
\noindent {\bf Vidal's theorem}. {\sl Every $K$-partite state of $K$ quNits can be expressed 
in a separable basis as} 
\begin{equation} 
\Gamma^{i_1, \dots, i_K}=
\sum_{a_1=1}^{r_1} \sum_{a_2=1}^{r_2}\cdots 
\sum_{a_{K-1}=1}^{r_{K-1}} 
A_{a_1}^{i_1} A_{a_1a_2}^{i_2}
\cdots
 A_{a_{K-2} a_{K-1}}^{i_{K-1}}   A_{a_{K-1}}^{i_K} \ , 
\label{MPSK1ny}
\end{equation}

\noindent {\sl where, if the state is normalized,}
\begin{equation} \sum_{i_k=0}^{N-1}(A^{i_k})^\dagger A^{i_k} = {\mathbbm 1}_{r_k} \ , 
\label{Vidal2} \end{equation}
\begin{equation} \sum_{i_k =0}^{N-1}A^{i_k}\Lambda_k(A^{i_k})^\dagger = \Lambda_{k-1} \ , 
\label{Vidal3} \end{equation}

\noindent {\sl $\Lambda_k$ is a reduced density matrix in diagonal form, and the Schmidt ranks 
are bounded by} $r_k \leq N^{[K/2]}$.
\smallskip 

\noindent The largest Schmidt ranks occur in the middle of 
the chain; $[N/2]$ denotes the largest integer not larger than $N/2$. Conditions 
(\ref{Vidal2}-\ref{Vidal3}) fix the representation uniquely up to orderings of 
the Schmidt vector and possible degeneracies there.

We have chosen to 
begin this story with a theorem by 
Vidal (2003) \cite{Vi03}, which gives a canonical form for Matrix Product States. 
But the story itself is much older. It really began with Affleck, Kennedy, Lieb, and Tasaki 
(1987) \cite{AKLT87}, who studied the ground states of isotropic quantum antiferromagnets. 
This was followed up by Fannes, Nachtergaele and Werner (1992) \cite{FNW92}.
Important precursors include works by Baxter (1968) \cite{Ba68} and Accardi 
(1981)  \cite{Acc81}. 

Of course, once we have a state expressed in the form (\ref{MPSK1ny}) we can relax 
conditions (\ref{Vidal2}-\ref{Vidal3}) without changing the state---they simply 
offer a canonical form for the representation. An important 
variation of the theme must be mentioned. Eq. (\ref{MPSK1ny}) is 
said to use {\it open boundary conditions}, but one can also use {\it periodic 
boundary conditions}, which here means that one uses a collection of matrices 
such that 
\begin{equation} 
\Gamma^{i_1, \dots, i_K}= \mbox{Tr} A^{i_1} A^{i_2} \cdots A^{i_{K-1}}A^{i_K} \ .   
\label{MPSper} \end{equation}

\noindent This time there are no vectors at the ends, and indeed there are no ends---the 
chain is closed. Physically this formulation is often preferred, but there is no longer 
an obvious way to impose a canonical form on the matrices. 
 
\begin{center}
\begin{figure}[ht]
\includegraphics[width=100mm]{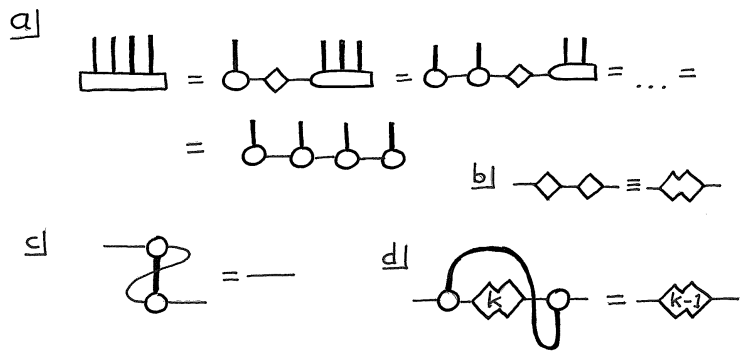}
\caption{Graphical notation for Matrix Product States. 
First recall Figure \ref{fig:birdtracks}. 
The equations that have been translated into graphical notation are: a)  
(\ref{firststep}), (\ref{secondstep}), and (\ref{MPSK1ny}), b) (\ref{Vidb}), c) 
(\ref{Vidal2}), and d) (\ref{Vidal3}). We try to make the lines proceed 
horizontally when the indices pertain to ${\mathbbm C}^D$ (so 
`arms and legs' are replaced by `left and right arms'). 
Lines pertaining to ${\mathbbm C}^N$ are drawn just a little bit thicker (confusion is 
unlikely to occur). 
}
\label{fig:MPSdecny}
\end{figure}
\end{center}

The only problem is that we risk getting lost in a clutter of indices. This is where 
the graphical notation for tensors comes into its own, see Figure \ref{fig:MPSdecny}. 

Having organized the $N^K$ components of the state into products of a collection of 
$KN$ matrices 
we must ask: was this a useful thing to do? Indeed yes, for two reasons. The first 
reason is that we are not interested in generic states, in 
fact our aim is to find special states obeying the Area Law. For this reason we 
now focus on states that can be written on the form (\ref{MPSK1ny}), but with all 
the Schmidt ranks obeying $r_k \leq D$, where $D$ is some modest integer called the 
{\it bond dimension}. We call them {\it matrix product states}, or {\it MPS} for 
short. 
Indispensable references, in addition to those mentioned above 
include papers by Perez-Garcia et al. \cite{PGVWC07} and 
by Verstraete, Murg, and Cirac \cite{VMC08}.
Splitting the chain in two and tracing out the 
contribution from one of the regions will result in a mixed state whose entanglement entropy behaves as 
\begin{equation} S(\rho_X) \sim \ln{D} \ . \end{equation}
 
\noindent We conclude that the strict Area Law holds for a one dimensional chain if and only 
if $D$ is independent of the number of subsystems. 

There are many interesting states with low bond dimensions. Separable states have 
bond dimension one. The GHZ$_K$ state has bond dimension two, and moreover the 
matrices $A^{i_k}$ for 
all sites $k$ between 2 and $K-1$ are the same, namely 
\begin{equation} A^0 = \left( \begin{array}{cc} 1 & 0 \\ 0 & 0 \end{array} \right) \ , 
\hspace{8mm} A^1 = \left( \begin{array}{cc} 0 & 0 \\ 0 & 1 \end{array} \right) \ . \end{equation}

\noindent With a suitable choice of vectors at the ends we recover the GHZ state. 
The $W_K$ state also has bond dimension two, but the bond dimension increases for 
the other Dicke states. Importantly, ground states of interesting Hamiltonians (leading 
to the Area Law) will have modest bond dimension, in particular the AKLT ground state 
\cite{AKLT87} (which started off the subject \cite{FNW92}) has bond dimension two. See Problem  17.12   
for some examples. 

The special properties of the singular value decomposition ensure that we have a 
reasonable approximation scheme 
\index{singular values}
on our hands. The {\it Eckart-Young theorem} says that if we want to approximate a matrix 
$M$ with another  matrix $\hat{M}$ of lower rank $D$, in such a way that the $L_2$-norm 
Tr$(\hat{M} - M)^\dagger (\hat{M} - M)$ is the smallest possible, then we can do this 
by performing the singular values decomposition of $M$ and 
setting all but the $D$ largest Schmidt coefficients to zero.
(The Eckart-Young 
theorem (1936) \cite{EY36} was later extended to cover all the $L_p$-norms.) 
This means that we can approximate any state by setting all but the $D$ largest Schmidt 
coefficients to zero (and renormalizing the remaining Schmidt coefficients so that 
they again sum to one), at each step of the exact expression (\ref{MPSK1ny}). The 
number of parameters in the approximation scales as $KND^2$, that is to say linearly 
in the number of subsystems. 
For a given $D$ the set of all MPS is a subset of measure zero in the set of all states, 
but we can make the approximation of any state better by increasing $D$. Truncating 
at some reasonable integer $D$ in a computer calculation is not so very different 
from representing real numbers by rational numbers with reasonable denominators. 
It works fine in everyday calculations. Conversely, if an exponential growth 
in the bond dimensions is encountered then the calculation cannot be done 
efficiently on a classical computer \cite{Vi03}.

We still have to present at least some evidence that ground states of interesting 
physical systems can be well approximated by MPS.
(Much more can be said on the topic why and how MPS represent 
ground states faithfully \cite{VC06}. We take a short-cut here.) 
 The question is if they obey the Area Law. 
They do, as proved by Hastings \cite{Ha07}. The conditions imposed 
on the Hamiltonian are that 
\begin{equation} H = \sum_{k=1}^KH_{k,k+1} \ , \hspace{8mm} ||H_{k,k+1}||_\infty \leq J 
\ , \hspace{8mm} \Delta E > 0 \ , \label{hamcondmat} \end{equation}

\noindent for some $J$ and for some energy gap $\Delta E$ between the ground state 
and the first excited state. The first condition insists that only nearest neighbours 
are coupled, the second bounds the largest eigenvalue of each individual term. 
Interactions of finite range can be dealt with by grouping sites together into 
single sites, so the restriction to interactions between nearest neighbours is 
not all that severe.

Hastings' proof is beyond our scope, but one key ingredient must be mentioned 
because we need it to state the theorem. Consider two disjoint regions $X$ and $Y$ 
of the lattice. Take operators $A$ and $B$ supported in $X$ and $Y$, respectively. 
Hence $[A,B] = 0$. Time evolve $A$ with a Hamiltonian obeying conditions (\ref{hamcondmat}), 
\begin{equation} A(t) = e^{iHt}Ae^{-iHt} \ . \end{equation}
 
\noindent Then one finds \cite{LiRo72}
\index{theorem!Lieb--Robinson}

\smallskip
\noindent {\bf Lieb--Robinson theorem}. {\sl Under the conditions stated there exist 
constants $c$ and $a$ and a velocity $v$ such that}
\begin{equation} || [A(t),B]||_\infty \leq c ||A||_\infty ||B||_\infty e^{-a(d(X,Y) - 
v|t|)} \ . \end{equation}

\noindent The constant $a$ is adjustable and can be chosen to be large provided that 
$d(X,Y)/v|t|$ is sufficiently large. Again the proof is beyond us, but take note of 
the physical meaning: this is a rigorous statement saying that an effective `light cone' 
appears in the system. Up to an exponentially decaying tail, influences cannot propagate 
outwards from the region $X$ faster than the Lieb--Robinson velocity $v$, and there 
will be definite bounds on how fast entanglement can spread through the system under 
local interactions \cite{BHV06}.

Now we can state \cite{Ha07}
\index{theorem!Hastings' area law}
\index{law!area}

\smallskip
\noindent {\bf Hastings' Area Law theorem}. {\sl Let $X$ be the region to the left 
(or right) of any site on a one-dimensional chain. For the ground state of a 
Hamiltonian obeying the conditions stated there holds the Area Law}
\begin{equation} S(\rho_X) \; \le \; c_0 \xi \ln{\xi}\ln{N}2^{\xi \ln{N}} \ , \end{equation}

\noindent {\sl where $c_0$ is a numerical constant of order unity and} 
\begin{equation} \xi = \mbox{max}\left( \frac{12v}{\Delta E}, 6a\right) \ . \end{equation}

\noindent {\sl Here $\Delta E$ is the energy gap, $v$ is the Lieb--Robinson velocity, and 
the constant $a$ is the one that appears in the Lieb--Robinson theorem.}
\smallskip
 
The point is that the upper bound on the entanglement entropy depends 
on the dimensionality $N$ of the subsystems, and on the parameters of the model, 
but not on the number of subsystems within the region $X$. All the vagueness in 
our statement of the Area Law has disappeared, at the expense of some precise 
limitations on the Hamiltonians that we admit. 
In the case of {\it critical systems}, for which $\Delta E \to 0$,
one observes \cite{CC09,LR09,ECB10} a logarithmic dependence of entropy on volume, 
$S(\rho_X) \approx c_1\log |X| +c_2$. This is 
a much milder growth than one expects from a generic state obeying the Volume 
Law. 
For one-dimensional chains area laws can also be derived assuming 
exponential decay of correlations
in the system, without any assumptions about the energy gap
\cite{BH13}.

Once it is admitted that the Area Law holds for the ground states of physically 
interesting Hamiltonians, the task of finding these states is greatly simplified. 
A standard approach is to solve the Rayleigh-Ritz problem: the ground state of 
the Hamiltonian  $H$  is the vector that minimizes the {\it Rayleigh quotient}  
\begin{equation} 
\frac{\langle \psi |H|\psi \rangle}
{\langle \psi|\psi \rangle } \ . \label{Rayleigh} \end{equation}

\noindent But the set of all states grows exponentially with the number $K$ of 
subsystems, so the best that can be done in practice is to find the minimum over 
a suitable set of trial states. The calculations become feasible once we assume that 
$|\psi\rangle $ belongs to the set of all MPS with a bond dimension growing at most 
polynomially in $K$. This idea is at 
the root of {\it the density matrix renormalization group} (DMRG) method, which 
has proven immensely useful in the study of strongly correlated one-dimensional 
systems.
(The DMGR is due to White (1992) \cite{Wh92}. 
The link to MPS was forged by \"Ostlund and Rommer (1995) \cite{OR95}. For a 
review -- with useful calculational details of MPS --
 see Schollw\"ock \cite{Sch11}.)

Note that if the computer is to be able to evaluate the Rayleigh quotient 
(\ref{Rayleigh}), it has to be told how. The first aim is to evaluate the 
single number 
\begin{equation} \langle \psi |\psi\rangle = \sum_{i_1,i_2, \dots ,i_K} 
\mbox{Tr}A^{i_1}A^{i_2} \cdots A^{i_K} 
\mbox{Tr}A^*_{i_1}A^*_{i_2} \cdots A^*_{i_K} \ . \end{equation}

\begin{center}
\begin{figure}[ht]
\includegraphics[width=100mm]{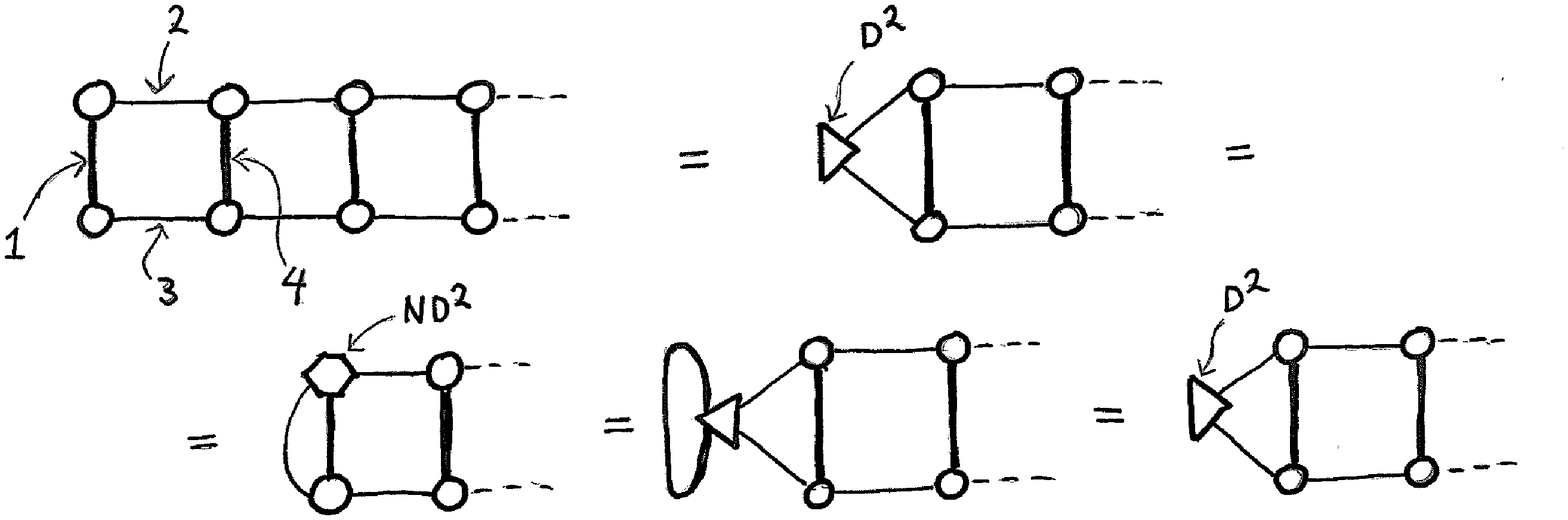}
\caption{Normalizing an MPS state: Eventually all the sums have to be done. Doing 
the contractions in the order indicated ensures that the number of components that 
have to be stored during intermediate stages of the calculation stays reasonable 
(as indicated by the arrows).
}
\label{fig:tensor1}
\end{figure}
\end{center}

\noindent It would clearly be a mistake to perform the traces first and the 
summation over the explicit indices afterwards---doing so means that the computer 
has to store $2N^K$ numbers at an intermediate stage of the calculation. In this 
case it is not difficult to propose a better strategy (and we do so in Figure 
\ref{fig:tensor1}) but for spin systems in spatial dimensions larger than one 
there are difficult issues of computational complexity to be addressed 
\cite{SWVC07}.
(To learn 
how to actually perform calculations the review by Or\'us \cite{Or14} and the 
paper by Huckle et al. \cite{HWSH13} may be helpful.)

So what about higher spatial dimension? Then the situation is not quite as clear-cut, 
but one still expects an area law to hold in suitable circumstances, even though 
the boundary $\partial X$ of a region $X$ in the lattice can have a complicated 
structure. In particular, with some assumptions about the decay of correlations and 
on the density of states, Masanes proved \cite{Ma09} that the entanglement entropy 
for a reduction of the ground state scales as
\begin{equation}
    S(\rho_X) \leq C |\partial X|(\ln{|X|})^N + O\left( |\partial X|(\ln{|X|}^{N-1}\right) 
		\ , 
\label{entaread}
 \end{equation}

\noindent where the constant $C$ depends on the parameters of the model, but not 
on the volume $|X|$. The ratio between the right hand side and the volume tends 
to zero as the size of the region grows, so although there is a logarithmic 
correction this again deserves to be called an Area Law. 

So it makes sense to look for a way to generalize MPS to higher dimensions. To this end 
we begin by looking at the one-dimensional construction in a different way. We begin 
by doubling each site in the chain, so that the total Hilbert space becomes 
${\cal H}_D^{\otimes 2K}$. The dimension of each factor is set to $D$, which may be 
larger than the original dimension $N$ of the physical subsystems. Then we consider 
a quite special state there, namely a product of maximally entangled bipartite states 
\begin{equation} |\phi^+_{k,k+1}\rangle = \sum_{c =0}^{D-1} 
|c \rangle_{k_R}|c \rangle_{(k+1)_L} \ . \end{equation}

\noindent (We worry about normalization only at the end of the construction.) 
The factor Hilbert spaces that occur here are the rightmost factor from site $k$ 
and the leftmost factor for site $k+1$. In effect entanglement is being used to 
link the sites together. The total state of the `virtual' (doubled) chain 
is taken to be the entangled pair state
\begin{equation} |\Psi \rangle = |\phi^+_{12}\rangle |\phi^+_{23}\rangle \dots 
|\phi^+_{K1}\rangle \ . \label{pepsfig1} \end{equation}   

\noindent Next, once for every site in the original chain, we introduce a linear 
map from the doubled Hilbert space ${\mathbbm C}^D\otimes {\mathbbm C}^D$ back to 
the $N$-dimensional Hilbert space ${\mathbbm C}^N$ we started out with:
\begin{equation} {\cal A}_k = A^{i_k}_{a_k b_k}|i_k\rangle 
\langle a_k b_k| \ , \end{equation}

\noindent where summation over repeated indices is understood. It is now a straightforward 
exercise to show that 
\begin{equation} {\cal A}_1{\cal A}_2 \dots {\cal A}_K|\Psi\rangle = 
A^{i_1}_{a_1a_2}A^{i_2}_{a_2a_3} \dots A^{i_K}_{a_Ka_1}
|i_ii_2 \dots i_K\rangle \ . \label{pepsfig2} \end{equation}

\begin{center}
\begin{figure}[ht]
\includegraphics[width=80mm]{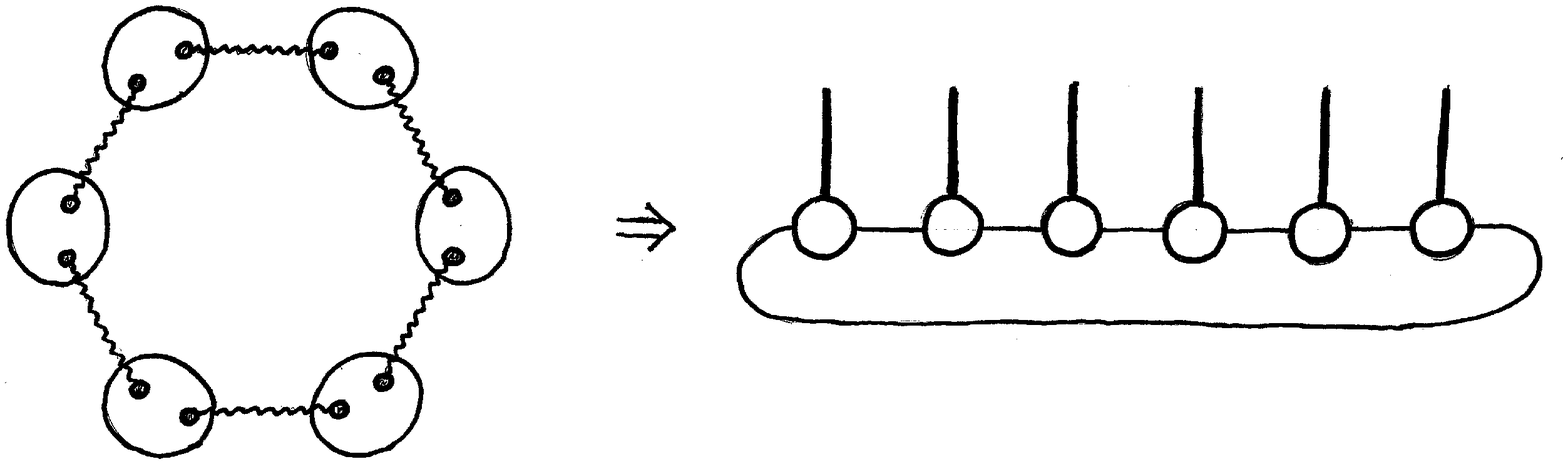}
\caption{The PEPS construction. We start from a state consisting of entangled pairs in an 
auxiliary Hilbert space (forming a ring in this example), apply a linear map, and obtain 
a Matrix Product State (with periodic boundary conditions).}
\label{fig:peps}
\end{figure}
\end{center}

\noindent We have recovered a Matrix Product State with periodic boundary conditions, as 
in Eq. (\ref{MPSper}). When arrived at in this way it is called a {\it projected 
entangled pair state}, abbreviated {\it PEPS} \cite{VC04}. 
\index{states!projected entangled pair}
See Figure \ref{fig:peps}. 

\begin{center}
\begin{figure}[ht]
\includegraphics[width=90mm]{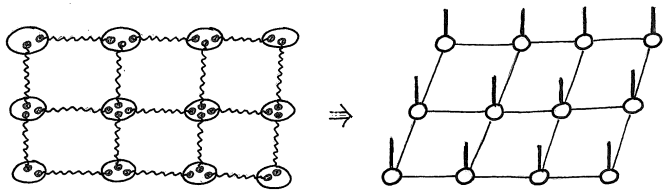}
\caption{The PEPS construction works in all spatial dimensions. Applying the map 
we find that the tensors in the interior of the lattice have $4 + 1$ arms. In this 
way we obtain a tensor network, rather than just a matrix product state.
}
\label{fig:tensor2}
\end{figure}
\end{center}

The PEPS construction is easily generalized to any spatial dimension of the lattice. 
As shown in Figure \ref{fig:tensor2}, to describe a two-dimensional lattice we 
have to expand the Hilbert space ${\mathbbm C}^N$ into a four--partite Hilbert space 
$({\mathbbm C}^N)^{\otimes 4}$, and then we introduce a perhaps site-dependent linear 
map $({\mathbbm C}^N)^{\otimes 4} \rightarrow {\mathbbm C}^N$. Clearly any lattice, in 
any spatial dimension, can be handled in a similar way. But going beyond one 
dimension does increase every calculational difficulty, and we break off the 
story here. More can be found in the references we have cited.  

We have now seen some simple examples of {\it tensor networks}.
\index{tensor network} 
The basic idea is to view a tensor 
of valence $m$ as an object with $m$ free legs, or with both arms and legs 
if the distinction between upper and lower indices is important, and perhaps 
with several different kinds of arms and legs. Then we choose an arbitrary 
undirected graph---this may be a one dimensional chain, a cubic lattice, or 
something much more general---and assign a tensor to each vertex, in such a 
way that each edge in the graph corresponds to a pair of contracted indices. 
Tensor networks have a long history and an active present. From the beginning 
there was a dream to see the geometry of space emerge from the geometry 
of quantum states \cite{Pe71}. 
Perhaps the application of tensor networks to quantum spin systems 
is beginning to substantiate the dream?

\medskip

\section{Concluding remarks}

\vspace{5mm}

The aim of these notes is literally to present
an concise introduction to the subject of
multipartite entanglement. 
%
%
We believe such a knowledge will contribute to a 
better understanding of quantum mechanics.
We hope also that it will provide a solid foundation
for various modern applications of quantum theory 
including quantum cryptography,
quantum error correction, and quantum computing.

There is much more to say. We have said nothing about graph states or toric codes, 
which is where the multipartite Heisenberg group comes into play. We have said 
nothing about the vast field of mixed multipartite states, nothing about multipartite 
Bell inequalities, and nothing about fermionic systems. 
The list of omissions can 
be made longer. But we have to stop somewhere.

\medskip 
           \centerline{ *  *  *  }   
\medskip 
 
Entanglement plays a crucial role in quantum information processing:
 it can be 
considered as our enemy or as our friend \cite{Pres98}.
On the negative side, the inevitable interaction systems with their environment  
induces entanglement between the controlled subsystems 
and the rest of the world,
which influences the state of the qubits and induces errors.
On the positive side, entanglement 
allows us to encode a single qubit in larger quantum systems, 
so that the entire quantum information will not be destroyed if 
the environment interacts with a small number of qubits. 
These issues become specially significant if one considers
systems consisting of several parties.

\medskip

It is tempting to compare quantum entanglement with the snow found 
high in the mountains during a late spring excursion. A mountaineer equipped 
with touring skis or crampones and iceaxe typically looks for couloirs and 
slopes covered by snow. On the other hand, his colleague in light climbing 
shoes will try to avoid all snowy fields to find safe passages across the rocks. 
The analogy holds as neither quantum entanglement nor spring snow lasts forever, 
even though the decay timescales do differ.

We wrap up these notes with the remark that multipartite entanglement
offers a lot of space for effects not present in the case of systems consisting of two subsystems only. 
Entanglement in many body systems is not well understood, 
so we are pleased to encourage the 
reader to contribute to this challenging field.

\bigskip


We are indebted to 
Rados{\l}aw Adamczak, 
Ole Andersson, 
Marcus Appleby, 
 Runyao Duan, Shmuel Friedland, 
Dardo Goyeneche, Marcus Grassl,
David Gross, Micha{\l} and Pawe{\l} Horodeccy, 
Ted Jacobson, Marek Ku\'s, Ion Nechita, 
Zbigniew Pucha{\l}a, Wojciech Roga, Adam Sawicki, 
Andreas Winter, 
Iwona Wintrowicz, and Huangjun Zhu, for reading some fragments 
of the text and providing us with valuable remarks. 
We thank Kate Blanchfield, 
Piotr Gawron, Lia Pugliese, Konrad Szyma{\'n}ski, 
and Maria {\.Z}yczkowska, for preparing for us two dimensional
figures and three dimensional printouts, 
models and photos for the book.

\smallskip

Financial support by Narodowe Centrum Nauki 
under the grant number DEC-2015/18/A/ST2/00274
is gratefully acknowledged.

\appendix

\medskip
\section{Contents of the II edition of the book "Geometry of Quantum States. 
         An Introduction to Quantum Entanglement"
         by {\sl I. Bengtsson and K. {\.Z}yczkowski}}

{\small

 {\bf 1 Convexity, colours and statistics }
 
\quad {1.1} Convex sets                                

\quad {1.2} High dimensional geometry

\quad  {1.3} Colour theory

\quad  {1.4} What is ``distance''? 

\quad  {1.5} Probability and statistics

{\bf 2  Geometry of probability distributions } 

\quad {2.1} Majorization and partial order

\quad {2.2} Shannon entropy

\quad {2.3} Relative entropy

 \quad {2.4} Continuous distributions and measures

\quad {2.5} Statistical geometry and the Fisher--Rao metric

 \quad {2.6} Classical ensembles

\quad {2.7} Generalized entropies

 {\bf 3 Much ado about spheres}
 
\quad {3.1} Spheres

\quad  {3.2} Parallel transport and statistical geometry

 \quad {3.3} Complex, Hermitian, and K\"ahler manifolds

\quad  {3.4} Symplectic manifolds

\quad {3.5} The Hopf fibration of the $3$-sphere

\quad  {3.6} Fibre bundles and their connections

\quad  {3.7} The $3$-sphere as a group

\quad  {3.8} Cosets and all that

 {\bf 4  Complex projective spaces}

\quad  {4.1} From art to mathematics

\quad  {4.2} Complex projective geometry

\quad  {4.3} Complex curves, quadrics and the Segre embedding

\quad {4.4} Stars, spinors, and complex curves

\quad {4.5} The Fubini-Study metric

\quad {4.6} ${\mathbb C}{\bf P}^n$ illustrated

\quad {4.7} Symplectic geometry and the Fubini--Study measure

\quad {4.8} Fibre bundle aspects

\quad {4.9} Grassmannians and flag manifolds

 {\bf 5 Outline of quantum mechanics}

 \quad{5.1} Quantum mechanics

 \quad{5.2} Qubits and Bloch spheres

 \quad{5.3} The statistical and the Fubini-Study distances

 \quad{5.4} A real look at quantum dynamics

 \quad{5.5} Time reversals

 \quad{5.6} Classical \& quantum states: a unified approach

 \quad{5.7} Gleason and Kochen-Specker 

 {\bf 6 Coherent states and group actions }

 \quad{6.1} Canonical coherent states

 \quad{6.2} Quasi-probability distributions on the plane

 \quad{6.3} Bloch coherent states

 \quad{6.4} From complex curves to $SU(K)$ coherent states

 \quad{6.5} $SU(3)$ coherent states

 {\bf 7 The stellar representation}

 \quad{7.1} The stellar representation in quantum mechanics

 \quad{7.2} Orbits and coherent states

 \quad{7.3} The Husimi function

 \quad{7.4} Wehrl entropy and the Lieb conjecture

 \quad{7.5} Generalised Wehrl entropies

 \quad{7.6} Random pure states

 \quad{7.7} From the transport problem to the Monge distance

 {\bf 8 The space of density matrices}

 \quad{8.1} Hilbert--Schmidt space and positive operators

 \quad{8.2} The set of mixed states

 \quad{8.3} Unitary transformations

 \quad{8.4} The space of density matrices as a convex set

\quad{8.5} Stratification

\quad{8.6} Projections and cross--sections 

 \quad{8.7} An algebraic afterthought

 \quad{8.8} Summary

 {\bf 9 Purification of mixed quantum states}

 \quad{9.1} Tensor products and state reduction

\quad{9.2} The Schmidt decomposition

\quad {9.3} State purification \& the Hilbert-Schmidt bundle

\quad {9.4} A first look at the Bures metric

 \quad{9.5} Bures geometry for $N=2$

 \quad{9.6} Further properties of the Bures metric

 {\bf 10 Quantum operations} 

 \quad{10.1} Measurements and POVMs

 \quad{10.2} Algebraic detour: matrix reshaping and reshuffling

 \quad{10.3} Positive and completely positive maps

\quad {10.4} Environmental representations

\quad {10.5} Some spectral properties

\quad{10.6} Unital \& bistochastic maps

\quad {10.7} One qubit maps

 {\bf 11 Duality: maps versus states}

 \quad{11.1} Positive \& decomposable maps

\quad {11.2} Dual cones and super-positive maps

 \quad{11.3} Jamio{\l }kowski isomorphism

 \quad{11.4} Quantum maps and quantum states

 {\bf 12  Discrete structures in Hilbert space} 

\quad{12.1}  Unitary operator bases and the Heisenberg groups

\quad{12.2} Prime, composite, and prime power dimensions

\quad{12.3}  More unitary operator bases 

\quad{12.4} Mutually unbiased bases

\quad{12.5} Finite geometries and discrete Wigner functions

\quad{12.6} Clifford groups and stabilizer states

\quad{12.7} Some designs

\quad{12.8} SICs

{\bf 13  Density matrices and entropies} 

 \quad{13.1} Ordering operators

 \quad{13.2} Von Neumann entropy

\quad {13.3} Quantum relative entropy

\quad {13.4} Other entropies

 \quad{13.5} Majorization of density matrices

\quad{13.6} Proof of the Lieb conjecture

\quad{13.7} Entropy dynamics

 {\bf 14 Distinguishability measures}

\quad{14.1} Classical distinguishability measures

\quad {14.2} Quantum distinguishability measures

\quad {14.3} Fidelity and statistical distance

 {\bf 15 Monotone metrics and measures}

\quad {15.1} Monotone metrics

\quad {15.2} Product measures and flag manifolds

 \quad{15.3} Hilbert-Schmidt measure

 \quad{15.4} Bures measure

 \quad{15.5} Induced measures

 \quad{15.6} Random density matrices

\quad {15.7} Random operations

\quad{15.8} Concentration of measure

 {\bf 16 Quantum entanglement}

 \quad{16.1} Introducing entanglement

 \quad{16.2} Two qubit pure states: entanglement illustrated

 \quad{16.3} Maximally entangled states

 \quad{16.4} Pure states of a bipartite system

\quad{16.5} A first look at entangled mixed states 

\quad{16.6} Separability criteria

\quad{16.7} Geometry of the set of separable states

 \quad{16.8} Entanglement measures

 \quad{16.9} Two qubit mixed states 

{\bf 17 Multipartite entanglement}

 \quad{17.1} How much is three larger than two? 

 \quad{17.2} Botany of states

\quad{17.3} Permutation symmetric states

\quad{17.4} Invariant theory and quantum states 

\quad{17.5} Monogamy relations and global multipartite entanglement

\quad{17.6} Local spectra and the momentum map

\quad{17.7} AME states and error--correcting codes 

\quad{17.8} Entanglement in quantum spin systems

{\bf Epilogue }

\quad  Appendix 1 Basic notions of differential geometry 

\quad Appendix 2 Basic notions of group theory 

\quad Appendix 3  Geometry do it yourself 

\quad Appendix 4  Hints and answers to the exercises
} 

\bigskip




\end{document}